\documentclass[12pt,a4paper]{article}

\usepackage{graphicx}
\usepackage{amsmath}
\usepackage{amssymb}
\usepackage[mathscr]{eucal}

\usepackage[pagebackref=false]{hyperref}
\hypersetup{
    colorlinks = true,
    extension = notused,  
    linkcolor = blue,
    anchorcolor = red,
    citecolor = blue,
    filecolor = blue,
    pagecolor = red,
    urlcolor = blue
}

\usepackage[height=8.8in, width=6.45in]{geometry}

\newcommand{\rch}{r}

\numberwithin{equation}{section}

\begin{document}
\begin{titlepage}

\begin{flushright}
YITP-14-110\\
\end{flushright}

\vskip 3cm

\begin{center}
{\Large \bfseries
A review on SUSY gauge theories on $\bf S^3$
}

\vskip 1.2cm

Kazuo Hosomichi

\bigskip
\bigskip

\begin{tabular}{ll}
& {\sl
 Department of Physics, National Taiwan University, Taipei 10617, Taiwan} \\
\end{tabular}
\vskip 1.5cm

\textbf{abstract}
\end{center}

\medskip
\noindent

We review the exact computations in 3D ${\cal N}=2$ supersymmetric gauge
theories on the round or squashed $S^3$ and the relation between 3D
partition functions and 4D superconformal indices.
This is part of a combined review on the recent developments of the
2d-4d relation, edited by J.~Teschner.

\bigskip
\vfill
\end{titlepage}

\setcounter{tocdepth}{2}

\tableofcontents

\section{Introduction}\label{S3:sec:Intro}

Localization principle has been a powerful tool in the study of
supersymmetric field theories which allows one to evaluate certain
SUSY-preserving quantities by explicit path integration.
It was first applied to 3D SUSY gauge theories on $S^3$ in
\cite{Kapustin:2009kz}, where a closed formula for partition function
and Wilson loop was obtained for a class of ${\cal N}\ge2$ superconformal
Chern-Simons matter theories. 
With generalization by \cite{Jafferis:2010un,Hama:2010av}, exact formula
is now available for arbitrary 3D ${\cal N}=2$ SUSY gauge theories.
The essential idea of localization is that, since nonzero contribution
to supersymmetric path integrals arise only from SUSY invariant configurations
of bosonic fields called saddle points, infinite dimensional path
integrals can be reduced to finite-dimensional integrals over saddle points.
It turned out that the analysis of 3D gauge theories on $S^3$ is much simpler
than the case of 4D ${\cal N}=2$ SUSY gauge theories on
$S^4$ \cite{Pestun:2007rz} (see \cite{P} for a review in this volume),
due to the absence of saddle points with non-trivial topological quantum
numbers.

The exact partition function, which depends on the radius of $S^3$ as
well as some of the coupling constants, is one of the most basic
quantities characterizing ${\cal N}=2$ supersymmetric theories.
More informaition about the theories can be obtained by putting them
on different 3D backgrounds preserving rigid supersymmetry and
evaluating partition functions. In \cite{Hama:2011ea} it was shown that
one can construct rigid ${\cal N}=2$ SUSY gauge theories on the
ellipsoid $S^3_b$ with $U(1)\times U(1)$ isometry,
\begin{equation}
 b^2(x_0^2+x_1^2)+b^{-2}(x_2^2+x_3^2)~=~1,
 \label{S3:eq:S3b}
\end{equation}
with a suitable background vector and scalar fields.
The additional fields which are required to make the ellipsoid
supersymmetric have their origin in the off-shell supergravity
\cite{Festuccia:2011ws}, where the fully generalized form of Killing
spinor equation appears as local SUSY transformation laws of fermions in
the supergravity multiplet.
The ellipsoid partition function was shown to depend on the squashing
parameter $b$ in a nontrivial manner.
Another important background with rigid supersymmetry is $S^2\times S^1$
which leads to the path integral definition of the 3D superconformal
index \cite{Kim:2009wb,Imamura:2011su}.
There are also results on more general 3D manifolds with a slightly
different formalism based on topological twist
\cite{Kallen:2011ny,Ohta:2012ev}.

Another useful approach to find supersymmetric deformations of the round
$S^3$ is the Scherk-Schwarz like reduction of $S^1\times S^3$, which means
that one includes finite rotation in the $S^3$ direction in the periodic
identification of fields along $S^1$. This approach also makes an
explicit connection between the 3D partition functions and 4D
superconformal indices
\cite{Romelsberger:2005eg,Kinney:2005ej,Romelsberger:2007ec,Dolan:2008qi},
and in particular the relation between nonzero angular momentum fugacity
in 4D and the deformed geometry in 3D \cite{Gadde:2011ia,Dolan:2011rp}.
As was shown in \cite{Imamura:2011uw,Imamura:2011wg}, the dimensional
reduction results in the familiar squashed $S^3$ with $SU(2)\times U(1)$
isometry, with some additional background fields turned on.
However, there are two inequivalent reductions whose effect on the 3D
physical quantities are totally different.

Meanwhile, the study of certain domain walls in 4D ${\cal N}=2$
superconformal gauge theories in connection with AGT relation led to
a conjecture that there is a precise agreement between quantities in 3D
gauge theories on $S^3$ and the reprerentation theory of Virasoro or W
algebras \cite{Drukker:2010jp,Hosomichi:2010vh}. In general,
compactification of a $(2,0)$ theory on a Riemann surface $\Sigma$ gives
rise to several different (Lagrangian) descriptions that are related to
one another by S-duality \cite{Gaiotto:2009we}. The S-duality domain
walls are defined by gluing two mutually S-dual theories along an
interface, and therefore have a natural connection to the elements of
the mapping class group or Moore-Seiberg groupoid operation acting on
conformal blocks. In this respect, it is important that the
squashing parameter $b$ corresponds to the Liouville or Toda coupling
constant. Indeed, one of the building blocks of the ellipsoid partition
function is the double-sine function $s_b(x)$, which in our context is most
conveniently defined as the zeta-regularized infinite product
\cite{Quine:1993aa}
\begin{equation}
 s_b(x)~=~ \prod_{m,n\in\mathbb{Z}_{\ge0}}
 \frac{mb+nb^{-1}+\frac Q2-ix}{mb+nb^{-1}+\frac Q2+ix}.\quad
\big(Q\equiv b+\frac1b\big)
\label{S3:eq:sbx}
\end{equation}
The same function appears in the structure constants of Liouville or
Toda CFTs with coupling $b$.

This review is organized as follows. In Section \ref{S3:sec:3dAGT} we
review the correspondence between a 3D gauge theory and 2D conformal field
theories in the canonical example of the S-duality domain wall
in ${\cal N}=2^\ast$ $SU(2)$ super Yang-Mills theory. In Section
\ref{S3:sec:3DPF} we review the localization computation for 3D gauge
theories on the round $S^3$ and the ellipsoid $S^3_b$, and summarize the
formulae for partition function as well as expectation values of
loop observables. In Section \ref{S3:sec:4DInd} we review the
path integral computation of 4D superconformal index, and see how the
squashed $S^3$ background arises as a result of Scherk-Schwarz
reduction.

\section{3D AGT relation}\label{S3:sec:3dAGT}

We review here the correspondence between 3D gauge theories and 2D
conformal field theories in one typical example. The original idea was
given in \cite{Drukker:2010jp} which discussed the S-duality domain walls
in 4D ${\cal N}=2$ superconformal theories of {\it class S}, namely the
compactification of $(2,0)$-theories on punctured Riemann surfaces
(see \cite{Ga} for a review).
It is important to recall here that, for this class of theories,
there are different gauge theory descriptions corresponding to different
pants decomposition $\sigma$ of the surface $\Sigma$, and they are
equivalent (S-dual) to one another. Also, if Lagrangian description is
available, its gauge coupling $q$ is determined from the complex
structure of $\Sigma$ which we regard to take values in Teichm\"uller space.

\subsection{Janus and S-duality domain walls}\label{S3:sec:walls}

A Janus domain wall is a supersymmeric deformation of gauge theories
which makes the complexified gauge coupling jump across the
wall. Consider a theory of class S on $S^4$ with a Janus wall along the
equator $S^3$ where the two (left and right) hemispheres with couplings
$q$ and $q'$ meet. The 4D partition function in the presence of the wall
should be given by
\begin{equation}
 Z ~=~ \int d\nu(a) \
 {\cal F}^{(\sigma)}_{a,m}(\bar q)
 {\cal F}^{(\sigma)}_{a,m}(q'),
\label{S3:eq:Zj}
\end{equation}
as the product of the instanton partition functions ${\cal F}$
integrated over the real Coulomb branch parameters $a$ with an appropriate
measure. Here $m$ denotes a collection of mass parameters, and $\sigma$
labels a choice of pants decomposition. For generic complex structure
$q$ there is a natural pants decomposition which leads to a weakly
coupled gauge theory description, and we choose $\sigma$ to be the
natural one at $q$.

As $q'$ is varied away from $q$, the gauge theory on the right hemisphere
becomes strongly coupled. To analytically continue the formula
(\ref{S3:eq:Zj}) in such a situation, one needs to S-dualize the right
hemisphere and move to another pants decomposition $\sigma'$ which gives
a weakly coupled description at $q'$. We then have a system of two
mutually S-dual theories meeting along the so-called S-duality domain wall.
In the special case where $q'$ is an image of $q$ under the mapping
class group, $\sigma$ and $\sigma'$ are equivalent so the
theories on the two sides of the wall are the same. However, their
degrees of freedom are connected across the wall via S-duality.

Under the AGT relation, the instanton partition functions correspond to
Liouville or Toda conformal blocks labeled by a fusion channel
$\sigma$ and the internal and external momenta $a,m$.
They should therefore transform under S-duality in the same way that the
corresponding conformal blocks transform under the Moore-Seiberg
groupoid operation $g$,
\begin{equation}
 {\cal F}_{a,m}^{(\sigma)}(q') ~=~
 \int d\nu(a')\,
 g_{a,a',m}\,
 {\cal F}_{a',m}^{(\sigma')}(q').
\label{S3:eq:MS}
\end{equation}
By substituting (\ref{S3:eq:MS}) into (\ref{S3:eq:Zj}) we obtain a formula
for the $S^4$ partition function in the presence of an S-duality domain wall.
Now the integration variables get doubled, as the Coulomb branch parameters
on the two sides of the wall can vary independently. At this point,
it is natural to expect that the integration kernel $g_{a,a',m}$
in (\ref{S3:eq:MS}) corresponds to the degrees of freedom localized on the
S-duality wall between the two 4D theories in their vacua $a,a'$.

In general, the S-duality walls should be described by some local 3D
worldvolume field theories coupled to the 4D bulk degrees of freedom.
In the following we take the example of ${\cal N}=2^\ast$ SYM
theory, which is a deformation of ${\cal N}=4$ SYM by a mass of the
adjoint hypermultiplet. The S-duality transformations for this theory
form the group $SL(2,\mathbb Z)$ and we are interested in the wall
corresponging to the ``$S$-element''. For $SU(N)$ gauge group, we expect
the correspondence with the $A_{N-1}$ Toda theory on a one-punctured
torus. In the Liouville case $N=2$, the kernel for the S-duality
operation acting on torus 1-point conformal block is known explicitly
\cite{Teschner:2003at},
\begin{equation}
 g_{(p,p',p_E)} = \frac{2^{\frac32}}{s_b(p_E)}
 \int_{\mathbb{R}}{\rm d}\sigma
 \frac{s_b(p'+\sigma+\frac12p_E+\frac{iQ}4)}
      {s_b(p'+\sigma-\frac12p_E-\frac{iQ}4)}
 \frac{s_b(p'-\sigma+\frac12p_E+\frac{iQ}4)}
      {s_b(p'-\sigma-\frac12p_E-\frac{iQ}4)}
 e^{4\pi ip\sigma}.
\label{S3:eq:sdk}
\end{equation}
Here $b$ is the Liouville coupling and $Q\equiv b+b^{-1}$. The
Liouville momenta $p,p',p_E$ are related to the conformal weight $h$
labeling the Virasoro highest weight representations by the formula
$h=p^2+Q^2/4$. The double-sine function $s_b(x)$ is defined by
(\ref{S3:eq:sbx}), and will appear frequently later in this article.

\subsection{Example: ${\cal N}=2^\ast$ SYM}\label{S3:sec:TSU2}

\begin{figure}[t]
\begin{center}
\raisebox{12mm}{\includegraphics[width=80mm]{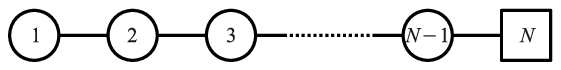}}
\raisebox{ 0mm}{\includegraphics[width=80mm]{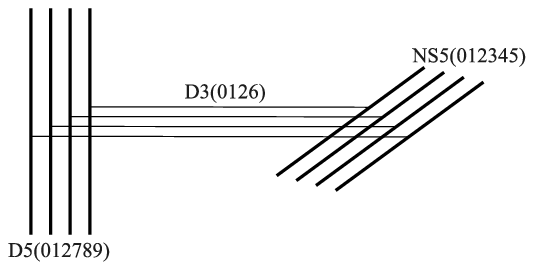}}
\end{center}
\vskip2mm
\caption{The quiver diagram and a type IIB brane construction for the 3D
 gauge theory $T[SU(N)]$.}
\label{S3:fig:tsun}
\end{figure}

A classification of boundary conditions and domain walls for
${\cal N}=4$ SYM theories with general gauge group $G$ was given in
\cite{Gaiotto:2008sa,Gaiotto:2008ak}, and the action of S-duality on
these objects was also studied. The 3D theory on the S-duality
domain walls, called $T[G]$, plays a central role in this story.
For $SU(N)$ gauge group, it was shown that the wall theory $T[SU(N)]$ is
given by a 3D ${\cal N}=4$ SUSY quiver gauge theory corresponding to the
diagram in the left of Figure \ref{S3:fig:tsun}.
Here the circles and the square correspond respectively to the gauge
symmetry $U(1)\times U(2)\times\cdots\times U(N-1)$ and a global $U(N)$
symmetry, and the links correspond to hypermultiplets.
The Coulomb and Higgs branch moduli spaces both have an $SU(N)$
symmetry which can be coupled to the gauge fields in the bulk.

A simple type IIB brane construction can reproduce this fact. Consider $N$
D3-branes stretched along the directions 0126 with $-L\le x_6\le L$,
ending on the D5-branes at $x_6=\pm L$ extending in the directions
012789. Due to the boundary condition at D5-branes, the massless
modes on D3-brane wordlvolume decompose into 3D ${\cal N}=4$ vector and
hypermultiplets. The vectormultiplet fields obey Dirichlet boundary
condition, so for small $L$ they are frozen to take vacuum
configuration. As was explained in \cite{Gaiotto:2008sa}, to avoid D3-branes
developing Nahm poles at the boundary, we need to introduce $N$ D5-branes
at each end so that each D5-brane has precisely one D3-brane ending on
it. Nonzero (real) Coulomb branch parameter $a$ can then be introduced
by putting the $i$-th D5-branes at, say, $(x_3,x_4,x_5)=(a_i,0,0)$ at each end.

Consider next the same brane configuration but now with an
S-duality domain wall on the D3-brane worldvolume at $x_6=0$.
It can be eliminated by applying the type IIB S-duality combined with
the exchange of 345 and 789 directions to the right half space $x_6\ge0$,
but then the $N$ D5-branes at $x_6=L$ turn into $N$ NS5-branes (012345).
The resulting brane configuration as shown on the right of Figure
\ref{S3:fig:tsun} is what precisely gives rise to the above-mentioned quiver
gauge theory. The D5-branes and NS5-branes are now free to move
independently. The positions of NS5-branes $a$ turn into $N-1$
Fayet-Iliopoulos parameters, whereas those of D5-branes $a'$ determine
the masses of the $U(N-1)\times U(N)$ bifundamental hypermultiplets.

Let us now focus on the simplest nontrivial case $N=2$. In 3D
${\cal N}=2$ terminology, the wall theory $T[SU(2)]$ is a $U(1)$ gauge
theory with five chiral multiplets $\phi,q_1,q_2,\tilde q^1,\tilde q^2$.
The neutral chiral field $\phi$ is a part of ${\cal N}=4$ $U(1)$ vector
multiplet and has R-charge 1. The two electrons $q_1,q_2$ and the two
positrons $\tilde q^1,\tilde q^2$ have the R-charge $1/2$, and they form
two flavors of hypermultiplets. ${\cal N}=4$ supersymmetry requires a
cubic superpotential of the form $\tilde q^i\phi q_i$.

As we have seen, the Coulomb branch parameter $a$ appears in
the wall theory as the $U(1)$ FI parameter, while $a'$ is the mass for
charged chiral fields which breaks the $SU(2)$ flavor symmetry to $U(1)$.
In addition, the bulk ${\cal N}=2^\ast$ mass parameter $m$ should also
show up in the wall theory in a way that preserves 3D ${\cal N}=2$
supersymmetry as well as the $SU(2)$ isometries of the Coulomb and Higgs
branches. It was argued in \cite{Hosomichi:2010vh} that $m$ is the mass
for the chiral fields associated to the global symmetry under which
$q_i,\tilde q^i$ have charge $+1$ and $\phi$ has charge $-2$.

It was observed in \cite{Hosomichi:2010vh} that the exact partition function
of this mass-deformed $T[SU(2)]$ theory on $S^3$ agrees precisely with
the kernel of the S-duality transformation (\ref{S3:eq:sdk}) for $b=1$,
under the identification
\begin{equation}
 p=a,\quad p'=a',\quad p_E=m.
\end{equation}
It was then shown in \cite{Hama:2011ea} that the formula
(\ref{S3:eq:sdk}) for general values of the coupling $b$ can be
reproduced by deforming the round $S^3$ into an ellipsoid $S^3_b$.
The derivation of the formulae which are necessary to confirm this
agreement will be reviewed in the next section.

\subsection{A 3D picture}\label{S3:sec:pict}

\begin{figure}[t]
\begin{center}
\includegraphics[width=0.3\hsize]{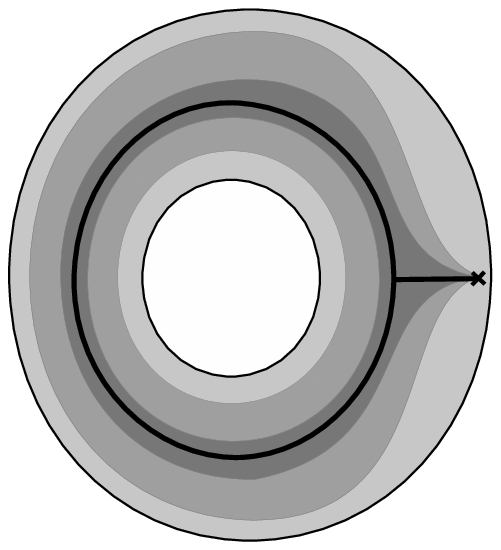}
\hskip10mm
\includegraphics[width=0.3\hsize]{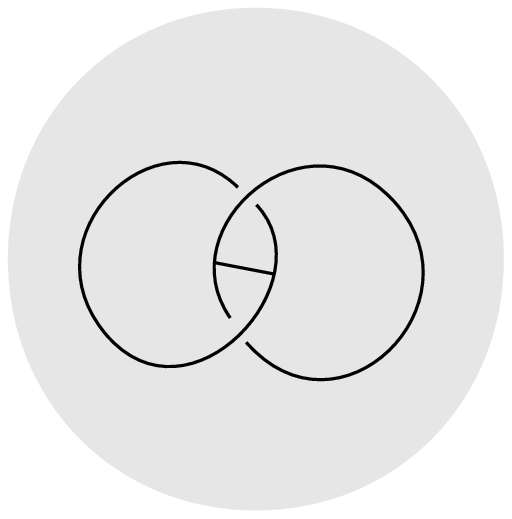}
\end{center}
\vskip2mm
\caption{(left) the process of a one-punctured torus degenerating into
 a Moore-Seiberg graph, thereby sweeping out a solid torus with a
 network of defect inside.
(right) Two such solid tori glued together to make an $S^3$ with a
 network of defect.}
\label{S3:fig:knot}
\end{figure}

As we have seen, Janus or S-duality domain walls correspond to smooth
evolutions of the complex structure of a surface, and therefore have an
interpretation as M5-branes wrapping three-manifolds. Let us explain
this in the example of ${\cal N}=2^\ast$ SYM.

Consider a Janus domain wall corresponding to a path in Teichm\"uller
space between two points of extreme weak coupling that are S-dual image
of each other. As one approaches towards one end from any point along the
path, the torus $\Sigma$ becomes thinner and thinner until it looks like
the Moore-Seiberg graph $\Gamma_1$ for the torus one-point conformal blocks.
In this process, two-dimensional part of the M5-brane worldvolume sweeps
out a 3D solid torus $B_1$ with a codimension-2 defect $\Gamma_1$ left
inside (Figure \ref{S3:fig:knot} left). One of the two basis 1-cycles
$\alpha,\beta$ of the torus, say $\alpha$, shrinks to zero length inside
$B_1$. Starting from the same point on the path and moving toward the
other end, one obtains another solid torus $B_2$ with a defect
$\Gamma_2$, inside which the cycle $\beta$ shrinks to zero length. The
two solid tori $B_1$ and $B_2$ glued together makes an $S^3$ with a
defect $\Gamma$ which is the union of the two graphs $\Gamma_1,\Gamma_2$
joined at the external legs (Figure \ref{S3:fig:knot} right). $\Gamma$
therefore consists of two circle defects and a segment connecting them,
and the three components are naturally labeled by the momenta $p,p', p_E$.

The 3D theories on domain walls or boundaries of 4D class S theories are
now regarded as part of a much bigger class of theories which arise from
M5-branes wrapping hyperbolic 3-manifolds. The relation between 3D SUSY
gauge theories and hyperbolic 3-manifolds also gives rise to an AGT-like
correspondence between 3D supersymmetric theories and Chern-Simons
theories with non-compact gauge groups. For more details on this topic,
see the review \cite{D} in this volume.

\section{3D Partition Function}\label{S3:sec:3DPF}

In this section we review the construction of 3D ${\cal N}=2$
supersymmetric gauge theories on a class of rigid SUSY backgrounds.
Then we concentrate on the theories on the round sphere and the
ellipsoids, and show how to compute partition function as well as the
expectation values of Wilson and vortex loops using localization principle.

\subsection{3D ${\cal N}=2$ SUSY theories}\label{S3:sec:3Dtheory}

Let us begin by summarizing our convention for 3D spinor calculus. We
use the standard Pauli's matrices for the Dirac matrices $\gamma^a$,
and also $\gamma^{ab}=\frac12(\gamma^a\gamma^b-\gamma^b\gamma^a)$.
To define bilinear products of spinors, we use an anti-symmetric $2\times2$
matrix $C$ with nonzero elements $C_{12}=-C_{21}=1$. Writing the spinor
indices explicitly, various bilinears are defined as follows.
\begin{equation}
 \epsilon\psi \equiv \epsilon^\alpha C_{\alpha\beta}\psi^\beta,
 \quad
 \epsilon\gamma^a\psi \equiv
 \epsilon^\alpha C_{\alpha\beta}(\gamma^a)^\beta_{~\gamma}\psi^\gamma,
 \quad
 \text{etc.} 
\end{equation}

In rigid SUSY theories on curved backgrounds, the parameters of SUSY
transformation $\epsilon$ are no longer constants, but are solutions to
the Killing spinor equation. For 3D ${\cal N}=2$ supersymmetric theories,
the SUSY is parametrized by two Killing spinors $\epsilon, \bar\epsilon$
of R-charge $+1, -1$. The most general form of the Killing spinor
equation can be found from off-shell supergravity \cite{Closset:2012ru}
as the condition that gravitini are invariant under local SUSY for a
suitable choice of parameters $\epsilon,\bar\epsilon$.
\begin{eqnarray}
 D_m\epsilon &=&
 \Big(\partial_m+\frac14\omega_m^{ab}\gamma^{ab}-iV_m\Big)\epsilon
 ~=~ iM\gamma_m\epsilon-iU_m\epsilon
 -\frac12\varepsilon_{mnp}U^n\gamma^p\epsilon,
 \nonumber \\
 D_m\bar\epsilon &=&
 \Big(\partial_m+\frac14\omega_m^{ab}\gamma^{ab}+iV_m\Big)\bar\epsilon
 ~=~ iM\gamma_m\bar\epsilon+iU_m\bar\epsilon
 +\frac12\varepsilon_{mnp}U^n\gamma^p\bar\epsilon.
\label{S3:eq:KS1}
\end{eqnarray}
Here $\gamma_m\equiv e^a_m\gamma^a$ with $e^a_m$ the vielbein, and
throughout this article we regard $\epsilon,\bar\epsilon$ as Grassmann even.
Supersymmetric backgrounds are therefore characterized by the metric
as well as the $U(1)_\text{R}$ gauge field $V_m$ and other auxiliary
fields $M,U_m$ in the off-shell gravity multiplet. In this section we
restrict our discussion to the backgrounds with
\begin{equation}
 U_m=0,
\label{S3:eq:KS1-2}
\end{equation}
which include the round sphere and ellipsoids. More general
supersymmetric backgrounds were studied systematically in three and four
dimensions in
\cite{Klare:2012gn,Dumitrescu:2012ha,Closset:2012ru,Alday:2013lba}.
For 3D ${\cal N}=2$ systems it was shown that the existence of a Killing
spinor implies that the background admits an almost contact metric structure.

The fields in 3D ${\cal N}=2$ theories are grouped into two kinds of
supermultiplets. A vector multiplet consists of a vector $A_m$, a real scalar
$\sigma$, a pair of spinors $\lambda,\bar\lambda$ and an auxiliary
scalar $D$ which are all Lie algebra valued. They transform under
supersymmetry as
\begin{eqnarray}
 \delta A_m &=&
 -\frac i2(\epsilon\gamma_m\bar\lambda+\bar\epsilon\gamma_m\lambda),
 \nonumber \\
 \delta\sigma &=&
 \frac12(\epsilon\bar\lambda-\bar\epsilon\lambda),
 \nonumber \\
 \delta\lambda &=&
 \frac12\gamma^{mn}\epsilon F_{mn}-\epsilon D
 -i\gamma^m\epsilon D_m\sigma,
 \nonumber \\
 \delta\bar\lambda &=&
 \frac12\gamma^{mn}\bar\epsilon F_{mn}+\bar\epsilon D
 +i\gamma^m\bar\epsilon D_m\sigma,
 \nonumber \\
 \delta D &=&
  \frac i2\epsilon\Big(\gamma^mD_m\bar\lambda+[\sigma,\bar\lambda]
  +iM\bar\lambda\Big)
 -\frac i2\bar\epsilon\Big(\gamma^mD_m\lambda-[\sigma,\lambda]
  +iM\lambda\Big).
\label{S3:eq:STv}
\end{eqnarray}
A chiral multiplet consists of a scalar $\phi$, a spinor $\psi$ and an
auxiliary scalar $F$ in an arbitrary representation $R$ of the gauge
group. Their conjugate fields $(\bar\phi,\bar\psi, \bar F)$ are in the
conjugate representation $\bar R$. If one assign the R-charge $\rch$ to
$\phi$ and $-\rch$ to $\bar\phi$, the R-charge of the remaining fields is
determined from the supersymmetry as in Table \ref{S3:tab:wr3}.
\begin{table}
\def\STRUT{\rule[-0.85mm]{0mm}{5.5mm}}
\begin{center}
\begin{tabular}{c||cc|ccccc|cccccc}
\hline
fields & $\epsilon$ & $\bar\epsilon$ &
 $A_a$ & $\sigma$ & $\lambda$ & $\bar\lambda$ & $\tilde D$ &
 $\phi$ & $\bar\phi$ & $\psi$ & $\bar\psi$ & $F$ & $\bar F$ \STRUT\\
\hline
weight & $-\frac12$ & $-\frac12$ &
 $1$ & $1$ & $\frac32$ & $\frac32$ & $2$ &
 $\rch$ & $\rch$ & $\rch+\frac12$ & $\rch+\frac12$ &
 $\rch+1$ & $\rch+1$ \STRUT\\
\hline
R-charge & $1$ & $-1$ &
 $0$ & $0$ & $1$ & $-1$ & $0$ &
 $\rch$ & $-\rch$ & $\rch-1$ & $1-\rch$ & $\rch-2$ & $2-\rch$ \STRUT\\
\hline
\end{tabular}
\caption{the scaling weight and the R-charge of the fields.}
\label{S3:tab:wr3}
\end{center}
\end{table}
The transformation rule for these fields is given by
\begin{eqnarray}
\delta\phi &=& \epsilon\psi,\quad
\delta\psi ~=~ i\gamma^m\bar\epsilon D_m\phi+i\bar\epsilon\sigma\phi
+\frac{2\rch i}3\gamma^mD_m\bar\epsilon\phi+\epsilon F,
\nonumber \\
\delta\bar\phi &=& \bar\epsilon\bar\psi,\quad
\delta\bar\psi ~=~
 i\gamma^m\epsilon D_m\bar\phi+i\epsilon\bar\phi\sigma
+\frac{2\rch i}3\gamma^mD_m\epsilon\bar\phi+\bar\epsilon\bar F,
\nonumber \\
\delta F &=&
\bar\epsilon(i\gamma^mD_m\psi-i\sigma\psi-i\bar\lambda\phi)
+\frac i3(2\rch-1)D_m\bar\epsilon\gamma^m\psi,
\nonumber \\
\delta\bar F &=&
 \epsilon(i\gamma^mD_m\bar\psi-i\bar\psi\sigma+i\bar\phi\lambda)
+\frac i3(2\rch-1)D_m\epsilon\gamma^m\bar\psi.
\end{eqnarray}
Here the quantities in the representation $R$ ($\bar R$) are regarded as
the column vectors (resp. row vectors), so that the vector multiplet
fields act on them from the left (right).

Supersymmetric Lagrangian consists of the following invariants. Those
involving only vector multiplet fields are the Chern-Simons term (for
which we write the action integral),
\begin{equation}
 S_\text{CS} ~=~ \frac{ik}{4\pi}\int
  \text{Tr}\bigg(A{\rm d}A-\frac{2i}3A^3
 -\sqrt{g}{\rm d}^3x
 \left(\bar\lambda\lambda+2\sigma D +4M\sigma^2\right)\bigg),
\end{equation}
the Yang-Mills term and the Fayet-Iliopoulos term for abelian gauge symmetry.
\begin{eqnarray}
 {\cal L}_\text{g} &=& 
 {\rm Tr}\bigg(\frac12F_{mn}F^{mn}+D_m\sigma D^m\sigma +D^2
 + i\bar\lambda\gamma^mD_m\lambda
 - i\bar\lambda[\sigma,\lambda]- M\bar\lambda\lambda\bigg),
 \nonumber \\
 {\cal L}_\text{FI} &=& -\frac{i\zeta}\pi \left(D+4M\sigma\right).
\end{eqnarray}
The kinetic term for chiral matters is given by
\begin{eqnarray}
 {\cal L}_\text{m} &=&
  D_m\bar\phi D^m\phi+\bar\phi\sigma^2\phi
 +4i(\rch-1)M\bar\phi\sigma\phi
 -2\rch(2\rch-1)M^2\bar\phi\phi
 +\frac{\rch R}4\bar\phi\phi-i\bar\phi D\phi
 \nonumber \\ &&
 +\bar FF
 -i\bar\psi\gamma^mD_m\psi+i\bar\psi\sigma\psi-(2\rch-1)M\bar\psi\psi
 +i\bar\psi\bar\lambda\phi-i\bar\phi\lambda\psi\,,
\end{eqnarray}
with $R$ the scalar curvature of the background.
The F-term of gauge invariant products of chiral multiplets with
R-charge $\rch=2$ is also invariant, but one can show that the result of
localization computation does not depend on the F-term couplings.
Note that, while the bosonic part of ${\cal L}_\text{g}$ is positive definite,
that of ${\cal L}_\text{m}$ has positive definite real part only when the
value of $\rch$ is chosen appropriately. For example, for round sphere
the positivity holds only when $0 <\rch< 2$.

The real mass for matters can be introduced by gauging the flavor
symmetry by a background vector multiplet.
The value of the background fields is chosen so as to preserve
supersymmetry,
\begin{equation}
 \sigma^\text{(bg)}=m~\text{(constant)},\quad
 D^\text{(bg)}=A_m^\text{(bg)}=\lambda^\text{(bg)}=\bar\lambda^\text{(bg)}=0.
\label{S3:eq:mass3}
\end{equation}

\subsection{SUSY localization}\label{S3:sec:localization}

To apply localization principle to supersymmetric path integrals, one
first chooses an arbitrary supercharge $\delta$, and then argue that the
nonzero contribution to the path integral can be localized to the vicinity of
saddle points, namely bosonic field configurations invariant under $\delta$.
This means that $\delta$-transform of all the fermions must vanish on
saddle points. For the theories of our interest, a useful observation is that
both ${\cal L}_\text{g}$ and ${\cal L}_\text{m}$ are SUSY exact for any
choice of $\delta$, which follows from
\begin{eqnarray}
 \bar\epsilon\epsilon\cdot
 {\cal L}_\text{g} &=& \delta_\epsilon\delta_{\bar\epsilon}
 {\rm Tr}\left(\bar\lambda\lambda+4D\sigma+8M\sigma^2 \right)\,,
 \nonumber \\
 \bar\epsilon\epsilon\cdot
 {\cal L}_\text{m} &=& \delta_\epsilon\delta_{\bar\epsilon}
 \left(\bar\psi\psi-2i\bar\phi\sigma\phi+4M(\rch-1)\bar\phi\phi\right)\,.
\end{eqnarray}
Namely, they can be written as $\delta$-variation of some fermionic
quantities, so they have to vanish at saddle points. A necessary
condition for vector multiplet fields at saddle points follows from
${\cal L}_\text{g}=0$,
\begin{equation}
 F_{mn}=D_m\sigma=D=0.
\label{S3:eq:sdl-vec}
\end{equation}
This is actually sufficient for the saddle point condition
$\delta\lambda=\delta\bar\lambda=0$ to be satisfied.
For theories on the round $S^3$ or its deformations, saddle
points are thus labeled by constant scalar field $\sigma$ and vanishing
gauge field, up to gauge transformations. For non-simply connected
manifolds such as lens spaces, one also has choices of Wilson lines
along non-contractible loops \cite{Gang:2009wy,Benini:2011nc,Imamura:2012rq}.
For matter multiplets, an obvious solution to $\delta\psi=\delta\bar\psi=0$ is
\begin{equation}
 \phi=\bar\phi=F=\bar F=0.
\label{S3:eq:sdl-ch}
\end{equation}
To show that this is the unique saddle point, the simplest way is to
check that the kinetic operator for $\phi$ in ${\cal L}_\text{m}$ has no
zeromodes, so that ${\cal L}_\text{m}$ vanishes only at (\ref{S3:eq:sdl-ch}).
For theories on the round sphere, one can show by a full spectrum analysis
that there are no zeromodes on all the saddle points as long as $0<\rch<2$.
This allows us to assume that the spectrum remains free of zeromodes on the
ellipsoids $S^3_b$ as long as $b$ is reasonably close to $1$.
The exact partition function on $S^3_b$ turns out to be analytic in $b$,
so it can be continued to arbitrary $b>0$.

Since ${\cal L}_\text{g}$ and ${\cal L}_\text{m}$ are exact, the
value of supersymmetric path integrals does not change if one adds them
to the original Lagrangian with arbitrary coefficients
$t_\text{g}, t_\text{m}$.
By making those coefficients very large, one can bring the theory into
extreme weak coupling. In this limit the path integral simplifies
and can be performed in two steps. One first integrates over
fluctuations around each saddle point, for which Gaussian approximation
is exact. The result is then integrated over the space of saddle points
labeled by constant $\sigma$.

\subsection{Partition function on the round sphere}\label{S3:sec:PFS3}

As the simplest and yet the most important case, let us reproduce
here the exact partition function of general ${\cal N}=2$ SUSY theories
on the unit round $S^3$.

We write the unit round metric as ${\rm d}s^2= e^ae^a$, and identify the
dreibein $e^a=e^a_m{\rm d}x^m$ with the left-invariant one-forms on the
$SU(2)$ group manifold via
\begin{equation}
 g^{-1}{\rm d}g=ie^a\gamma^a,\quad
 g\in SU(2).
\end{equation}
The isometry $SU(2)_\mathscr{L}\times SU(2)_\mathscr{R}$ acts on $g$ from
its left and right. Note that, under the above choice of the local
Lorentz frame, $SU(2)_\mathscr{R}$ acts on fields as local Lorentz
rotation as well as isometry rotation.

Let us summarize here the spectrum of free fields on the round
sphere. We first notice that one can use the inverse dreibein $e^{am}$
to define a triplet of vector fields
$\mathscr{R}^a\equiv\frac1{2i}e^{am}\partial_m$ which generates
$SU(2)_\mathscr{R}$. Using them, the kinetic terms for free complex
scalars and spinors can be rewritten as
\begin{eqnarray}
 \bar\phi\;\Delta_{S^3}^\text{scalar}\phi &\equiv&
 g^{mn}\partial_m\bar\phi\partial_n\phi ~=~
 \bar\phi\cdot 4\mathscr{R}^a\mathscr{R}^a\phi,
\nonumber \\
 -i\bar\psi\;\slash\!\!\!\! D_{S^3}\psi &\equiv&
 -i\bar\psi\gamma^mD_m\psi ~=~
 \bar\psi(4S^a\mathscr{R}^a+\tfrac32)\psi,
\end{eqnarray}
where $S^a=\frac12\gamma^a$ is the generator of local Lorentz $SU(2)$
acting on spinors. Likewise, for a free Maxwell field $A=A^ae^a$ and its
field strength $\ast {\rm d}A=F^ae^a$, one finds
\begin{equation}
 F^a~=~ 2i\varepsilon^{abc}\mathscr{R}^bA^c+2A^a,~~\text{or}\quad
 \vec F = (2+2\mathscr{R}^aT^a)\vec A,
\end{equation}
where $T^a$ is the generator of local Lorentz $SU(2)$ in the triplet
representation. The Maxwell kinetic operator for gauge field is
given by $\Delta_{S^3}^\text{vector}\equiv (\ast{\rm d})^2$. The space
of scalar, spinor and vector wave functions on $S^3$ thus form the following
representation of $SU(2)_\mathscr{L}\times SU(2)_\mathscr{R}$.
\begin{eqnarray}
 {\cal H}_\text{scalar} &=&
 \bigoplus_{n\ge0}(\tfrac n2,\tfrac n2)_{n(n+2)},
 \nonumber \\
 {\cal H}_\text{spinor} &=&
 \bigoplus_{n\ge0}\Big\{
 (\tfrac n2,\tfrac{n+1}2)_{n+3/2} \oplus
 (\tfrac{n+1}2,\tfrac n2)_{-n-3/2}
 \Big\},
 \nonumber \\
 {\cal H}_\text{vector} &=&
 \bigoplus_{n\ge0}\Big\{
 (\tfrac n2,\tfrac{n+2}2)_{(n+2)^2} \oplus
 (\tfrac{n+1}2,\tfrac{n+1}2)_{0} \oplus
 (\tfrac{n+2}2,\tfrac n2)_{(n+2)^2}
 \Big\}.
\end{eqnarray}
For convenience, we put the eigenvalue of
$\Delta_{S^3}^\text{scalar},-i\;\slash\!\!\!\! D_{S^3}$ or
$\Delta_{S^3}^\text{vector}$ for each irreducible representation as suffix.
Note that the nonzero eigenmodes of $\Delta_{S^3}^\text{vector}$ are
divergenceless vectors while the zero eigenmodes are total divergences.

On the unit round $S^3$, the simplest form of the Killing spinor equation
\begin{equation}
 \Big(\partial_m+\frac14\omega_m^{ab}\gamma^{ab}\Big)\epsilon ~=~
 iM\gamma_m\epsilon,\quad
 M=\pm\frac12
\label{S3:eq:KS-S3}
\end{equation}
has solutions. First, in the left-invariant local Lorentz frame, any
constant spinor satisfies (\ref{S3:eq:KS-S3}) with $M=+\frac12$. The two
independent solutions are left-invariant and transform as a doublet of
$SU(2)_{\mathscr{R}}$. In addition, there are two independent solutions
to (\ref{S3:eq:KS-S3}) with $M=-\frac12$ both of which are given by
$g^{-1}$ times a constant spinor. They are therefore right-invariant and form an
$SU(2)_{\mathscr{L}}$ doublet. In this subsection, we choose the background
$V_m=0, M=\frac12$.

Let us now turn to the computation of partition function using the
localization principle. The supersymmetric saddle points are labeled by
the constant value of the vector multiplet scalar $\sigma(x)=a$.
The Chern-Simons or Fayet-Iliopoulos Lagrangians take nonzero value at
the saddle point $a$ according to the formula
\begin{equation}
 e^{-S_\text{CS}}= e^{i\pi k\text{Tr}(a^2)}, \quad
 e^{-S_\text{FI}}= e^{4\pi i\zeta a},
\label{S3:eq:Scl}
\end{equation}
In addition, we need the one-loop determinant which arise from
integrating over all the fluctuation modes at the saddle point $a$ under
Gaussian (=one-loop) approximation.

We first study the vector multiplet for a non-abelian gauge
symmetry $G$. Following the general prescription, we add to the original
Lagrangian a SUSY exact regulator term $t_\text{g}{\cal L}_\text{g}$
and take $t_\text{g}\to\infty$. In this limit the regulator term
dominates the path integral weight, and the Gaussian approximation
becomes exact. The quadratic part of ${\cal L}_\text{g}$ in the Lorentz
gauge $\partial^mA_m=0$ is
\begin{equation}
 {\cal L}_\text{g} ~=~ {\rm Tr}\bigg[
  \vec A\big(\Delta_{S^3}^\text{vector}+a_\text{adj}^2\big)\vec A
 +\hat\sigma \Delta_{S^3}^\text{scalar}\hat\sigma
 +D^2
 -\bar\lambda\big(-\!i\;\slash\!\!\!\! D_{S^3}+\tfrac12
  +ia_\text{adj}\big)\lambda\bigg].
\end{equation}
Here we introduced the notation $a_\text{adj}$ for $a$ in the adjoint
representation, namely $a_\text{adj}\lambda\equiv [a,\lambda]$,
and $\hat\sigma$ denotes the fluctuation of $\sigma$ around its
saddle point value $a$.
To fix the gauge, we express the gauge field $A$ as a sum of a
divergenceless vector field $\hat A$ and a total derivative
${\rm d}\varphi$, and insert the delta functional for $\varphi$.
The Faddeev-Popov ghost determinant is trivial since gauge symmetry is
just the shift of $\varphi$ (up to terms irrelevant in the saddle-point
approximation). But since
${\rm Tr}A_mA^m={\rm Tr}
 (\hat A_m\hat A^m+\partial_m\varphi\partial^m\varphi)$, 
this change of integration variables gives rise to a Jacobian
\begin{equation}
 \mathscr{D}A~=~ \mathscr{D}\hat A\mathscr{D}'\varphi\cdot
 (\text{Det}'\Delta_{S^3}^\text{scalar})^{\frac12{\rm dim}G},
\end{equation}
where the primes indicate that the constant modes are excluded. This
Jacobian is canceled against the determinant arising from
$\hat\sigma$-integration.

The integration over the remaining physical fields $\lambda,\bar\lambda$
and $\hat A$ gives rise to the following ratio of determinants,
\begin{eqnarray}
 Z^\text{1-loop}_\text{vec} &=&
 \frac{\text{det}_\lambda(-ia-\frac12+i\;\slash\!\!\!\! D_{S^3})}
      {\text{det}_{\hat A}(a^2+\Delta_{S^3}^\text{vector})^{\frac12}}
 \nonumber \\ &=&
 \prod_{n\ge0}
 \frac{[\text{det}_\text{adj}(-ia-n-2)]^{(n+1)(n+2)}\cdot
       [\text{det}_\text{adj}(-ia+n+1)]^{(n+1)(n+2)}}
      {[\text{det}_\text{adj}(a^2+(n+2)^2)]^{(n+1)(n+3)}}\;.
\label{S3:eq:Zv1}
\end{eqnarray}
Let us take the Cartan-Weyl basis of $G$ and assume that the saddle point
parameter takes values in the Cargan subalgebra, namely $a=a_iH_i$ with
$H_i$ Cartan generators satisfying $\text{Tr}(H_iH_j)=\delta_{ij}$. The
above expression can then be rewritten further,
\begin{equation}
 Z^\text{1-loop}_\text{vec}
 ~=~
 \prod_{n\ge1}n^{2\text{rk}G}
 \prod_{\alpha\in\Delta_+}(n^2+(a\cdot\alpha)^2)^2
 ~=~
  (2\pi)^{\text{rk}G}\prod_{\alpha\in\Delta_+}
 \Big(\frac{2\sinh(\pi a\cdot\alpha)}{a\cdot\alpha}\Big)^2,
\label{S3:eq:Zv2}
\end{equation}
where $\alpha$ runs over all the positive roots. The divergent infinite
products were evaluated using zeta function regularization.

The constant value $a$ of the scalar field can always be gauge-rotated
into Cartan subalgebra. The domain of integration can therefore be
reduced to Cartan subalgebra, but this in turn introduces a
Vandermonde determinant in the measure which cancel nicely with the
denominator of (\ref{S3:eq:Zv2}). The exact partition function for a theory
with $G$ vector multiplet is thus an integral over its Cartan
subalgebra with the measure
\begin{equation}
 \frac1{|{\cal W}|}
 \prod_i{\rm d}a_i\;\prod_{\alpha\in\Delta_+}
 \big(2\sinh(\pi a\cdot\alpha)\big)^2.
\label{S3:eq:IM}
\end{equation}
Here we modded out by the order of the Weyl group ${\cal W}$, which is
the residual gauge symmetry after $a$ has been gauge rotated into
Cartan subalgebra.

Let us next turn to the matter fields. In the weak coupling limit, the
action ${\cal L}_\text{m}$ for the matter fluctuations at the saddle
point $a$ is given by
\begin{equation}
 {\cal L}_\text{m} =
 \bar\phi\{\Delta_{S^3}^\text{scalar}+a^2+2i(\rch-1)a+\rch(2-\rch)\}\phi
 +\bar FF
 +\bar\psi\{-i\;\slash\!\!\!\! D_{S^3}+\frac12+ia-\rch\}\psi\,.
\end{equation}
Let us choose the basis vectors $\{|w\rangle\}$ of the matter
representation $R$ so as to diagonalize Cartan generators,
i.e. $H_i|w\rangle=w_i|w\rangle$. Then the matter one-loop determinant
becomes,
\begin{eqnarray}
Z^\text{1-loop}_\text{matter} &=&
 \frac{\text{det}_\psi(\frac12+ia-\rch-i\;\slash\!\!\!\! D_{S^3})}
      {\text{det}_\phi(\Delta_{S^3}^\text{scalar}+1-(\rch-1-ia)^2)}
 \nonumber \\ &=&
 \prod_{n\ge0}\frac
 {[\text{det}_R(n+2+ia-\rch)]^{(n+1)(n+2)}\cdot
  [\text{det}_R(-n-1+ia-\rch)]^{(n+1)(n+2)}}
 {[\text{det}_R((n+1)^2-(\rch-1-ia)^2)]^{(n+1)^2}}
 \nonumber \\
 &=&
\prod_{n=1}^\infty\prod_w
\frac
{(n+1-\rch+ia\cdot w)^n}
{(n-1+\rch-ia\cdot w)^n}
 ~=~ \prod_w s_{b=1}(i(1-\rch)-a\cdot w),
\label{S3:eq:Zm1}
\end{eqnarray}
where $w$ runs over all the weights of $R$.

We thus arrived at an integral formula for exact partition function of
general 3D ${\cal N}=2$ SUSY gauge theories on the unit round
sphere. The basic building blocks for the integrand are the classical
action evaluated at saddle points (\ref{S3:eq:Scl}) and the matter one-loop
determinant (\ref{S3:eq:Zm1}), and their product is integrated over the
Cartan subalgebra of the gauge symmetry with the measure (\ref{S3:eq:IM}).
For theories with matter mass, the mass parameter $m$ of (\ref{S3:eq:mass3})
enters into the one-loop determinant (\ref{S3:eq:Zm1}) in the same way as
$a$, but we do not integrate over it.

\subsection{Partition function on ellipsoids}\label{S3:sec:PFS3b}

Let us next consider the deformation from the round sphere to
ellipsoids $S^3_b$ defined by (\ref{S3:eq:S3b}). With a suitable polar
coordinate system, the metric can be written as
\begin{eqnarray}
 {\rm d}s^2 &=&
 \frac1{b^2}\cos^2\theta {\rm d}\varphi^2
+b^2   \sin^2\theta {\rm d}\chi^2
+f^2{\rm d}\theta^2,
 \nonumber \\ &&
 f(\theta) ~=~ \sqrt{b^{-2}\sin^2\theta+b^2\cos^2\theta}.
\label{S3:eq:S3bp}
\end{eqnarray}
A natural choice for the dreibein and the resulting spin connection are
\begin{eqnarray}
 &&
 e^1= \frac1b\cos\theta {\rm d}\varphi,\quad
 e^2= b     \sin\theta {\rm d}\chi,\quad
 e^3=f{\rm d}\theta,
 \nonumber \\
 &&
 \omega^{12}=0,\quad
 \omega^{13}=-\frac1{bf}\sin\theta{\rm d}\varphi,\quad
 \omega^{23}= \frac b{f}\cos\theta{\rm d}\chi.
\end{eqnarray}

The ellipsoid can be made supersymmetric by turning on a suitable
$U(1)_\text{R}$ gauge field in the background. This was found in
\cite{Hama:2011ea} rather heuristically by taking a pair of Killing
spinors on the (unit) round sphere with $V_m=U_m=0$ and $M=\frac12$,
\begin{equation}
 \epsilon = \frac1{\sqrt2}
 \left(\begin{array}{r}
 -e^{\frac i2(\chi-\varphi+\theta)} \\
  e^{\frac i2(\chi-\varphi-\theta)} \end{array}\right),\quad
 \bar\epsilon = \frac1{\sqrt2}
 \left(\begin{array}{r}
  e^{\frac i2(-\chi+\varphi+\theta)} \\
  e^{\frac i2(-\chi+\varphi-\theta)} \end{array}\right),
\label{S3:eq:KSS3b}
\end{equation}
and studying the effect of squashing the metric. On the ellipsoid
(\ref{S3:eq:S3bp}) they were found to satisfy the Killing spinor equation
(\ref{S3:eq:KS1}) with $U_m=0$ and
\begin{equation}
 V ~=~ -\frac12\Big(1-\frac1{bf}\Big){\rm d}\varphi
       +\frac12\Big(1-\frac bf  \Big){\rm d}\chi,
\quad
 M~=~ \frac1{2f}.
\end{equation}
The supersymmetric observables on this background depend on the
squashing parameter $b$ in an nontrivial manner. Similar supersymmetric
deformations from the round $D$-sphere into ellipsoids were studied for
4D ${\cal N}=2$ theories by \cite{Hama:2012bg} and for 2D
${\cal N}=(2,2)$ theories by \cite{Gomis:2012wy}.

Note that, in finding the dreibein, spin connection and background
fields, the precise form of the function $f$ is actually not needed as
long as it is independent of $\varphi$ and $\chi$. It was pointed out in
\cite{Martelli:2011fu} that the above construction works for arbitrary
smooth $f(\theta)$, with the only requirement coming from the smoothness
at $\theta=0$ and $\frac\pi2$,
\begin{equation}
 f(\theta=0)=b,\quad f(\theta=\frac\pi2)=\frac1b.
\end{equation}
More general supersymmetric backgrounds of sphere topology was studied
in \cite{Alday:2013lba,Closset:2013vra}, but it was also shown that
supersymmetric observables depend on the background only through a
single parameter $b$. See also \cite{Nian:2013qwa,Tanaka:2013dca}.

The partition function on the ellipsoid background can be computed again
by applying the localization principle. First, the saddle points are
given by the solutions to (\ref{S3:eq:sdl-vec}) and (\ref{S3:eq:sdl-ch}) as for
the round sphere, and are therefore labeled by the constant value of the
vector multiplet scalar $\sigma$. The value of the CS and FI actions
$S_\text{CS}, S_\text{FI}$ also remain the same as (\ref{S3:eq:Scl}).
However, the evaluation of the one-loop determinants on the
ellipsoids (\ref{S3:eq:S3bp}) or other backgrounds with more general
$f(\theta)$ becomes more complicated since one can no longer work out
the full spectrum using spherical harmonics.

An alternative approach to compute the one-loop determinants is to study
how the supersymmetry relates bosonic and fermionic eigenmodes of the
Laplace or Dirac operators. Most of the eigenmodes are paired by the
supersymmetry so that their net contribution to the one-loop determinant
is trivial. It is therefore important to know the spectrum of the
eigenmodes without superpartner.

Let us begin with a chiral multiplet in a representation $R$ of the
gauge group $G$. We first move to a new set of fields
in terms of which the cancellation between bosonic and fermionic
eigenvalues is most transparent.
Let us introduce the Grassmann-odd scalar functions
$\Psi,\bar\Psi,\Psi',\bar\Psi'$ and Grassmann-even scalars $F',\bar F'$ by
\begin{eqnarray}
 \psi = \epsilon\Psi' -\bar\epsilon\Psi, &&
 F = F' -i\bar\epsilon\gamma^m\bar\epsilon D_m\phi,
 \nonumber \\
 \bar\psi = \bar\epsilon\bar\Psi' + \epsilon\bar\Psi,&&
 \bar F = \bar F' +i\epsilon\gamma^m\epsilon D_m\bar\phi.
\end{eqnarray}
They transform under supersymmetry as follows,
\begin{equation}
\begin{array}{l}
 \delta\phi=\Psi,\\
 \delta\bar\phi=\bar\Psi,
\end{array}
~~
\begin{array}{l}
 \delta\Psi=\mathscr{H}\phi,\\
 \delta\bar\Psi=\mathscr{H}\bar\phi,
\end{array}
~~
\begin{array}{l}
 \delta\Psi'=F',\\
 \delta\bar\Psi'=\bar F',
\end{array}
~~
\begin{array}{l}
 \delta F'= \mathscr{H}\Psi',\\
 \delta\bar F'=\mathscr{H}\bar\Psi',
\end{array}
\end{equation}
where $\mathscr{H}$ is the square of SUSY acting on scalar functions.
To be more explicit, it acts on $\phi$ carrying the R-charge $\rch$ as follows.
\begin{eqnarray}
 \mathscr{H}\phi
 &=& i\bar\epsilon\gamma^m\epsilon D_m\phi-i\sigma \phi+\frac\rch f\phi
 \nonumber \\ &=&
 \left\{-ib\partial_\varphi+ib^{-1}\partial_\chi-ia+\frac{Q\rch}2
 \right\}\phi\,.\quad
 \left(Q\equiv b+b^{-1}\right)
\end{eqnarray}
Here the second equality holds up to non-linear terms which are
irrelevant in the saddle point analysis.

To compute the one-loop determinant, we add a SUSY exact regulator
${\cal L}_\text{reg}=\delta{\cal V}$ to the original Lagrangian with a
large coefficient. We choose
\begin{equation}
 2{\cal V}
 = (\bar\phi,\bar F')\left(\begin{array}{cc}
  \mathscr{D} & 2\mathscr{D}_+ \\
  0 & 1 \end{array}\right)
 \left(\begin{array}{l} \Psi \\ \Psi' \end{array}\right)
 -(\bar\Psi,\bar\Psi')\left(\begin{array}{cc}
  \mathscr{D} & 0 \\
 2\mathscr{D}_- & -1 \end{array}\right)
 \left(\begin{array}{c} \phi \\ F' \end{array}\right),
\end{equation}
where
\begin{equation}
 \mathscr{D}  \equiv\mathscr{H}+2ia,\quad
 \mathscr{D}_+\equiv -i\epsilon\gamma^m\epsilon D_m,\quad
 \mathscr{D}_-\equiv  i\bar\epsilon\gamma^m\bar\epsilon D_m.
\end{equation}
One can show that the operators $\mathscr{D}_\pm$ commutes with
$\mathscr{H}$ by taking their R-charges $\pm2$ into account correctly.
The regulator Lagrangian ${\cal L}_\text{reg}$ in the quadratic
approximation consists of the following terms,
\begin{eqnarray}
 {\cal L}_\text{reg}\big|_\text{F}
 &=&
  (\bar\Psi,\bar\Psi')\left(\begin{array}{lc}
   \mathscr{D}   &  \mathscr{D}_+ \\
   \mathscr{D}_- & -\mathscr{H} \end{array}\right)
 \left(\begin{array}{c} \Psi \\ \Psi' \end{array}\right),
 \nonumber \\
 {\cal L}_\text{reg}|_\text{B}
 &=& (\bar\phi,\bar F')\left(\begin{array}{cc}
  \mathscr{D}\mathscr{H} & \mathscr{D}_+ \\
 -\mathscr{D}_-            & 1 \end{array}\right)
 \left(\begin{array}{c} \phi \\ F' \end{array}\right)
 ~=~\bar FF + \bar\phi\Delta\phi\,,
 \nonumber \\ &&
 \Delta = \mathscr{D}\mathscr{H}+\mathscr{D}_+\mathscr{D}_-
 = a^2 - (\bar\epsilon\gamma^m\epsilon D_m)^2
 + \epsilon\gamma^m\epsilon D_m\cdot\bar\epsilon\gamma^n\bar\epsilon D_n\,.
\label{S3:eq:Lreg}
\end{eqnarray}
Note that the bosonic part is positive definite.
Thus the one-loop determinant is given by the ratio of the determinants
for the Dirac operator (the $2\times 2$ matrix in the first line of
(\ref{S3:eq:Lreg})) and the Laplace operator $\Delta$.

As was shown in \cite{Hama:2011ea}, generically a scalar eigenmode of
$\Delta$ and a pair of Dirac eigenmodes form a multiplet which yields no net
contribution to the one-loop determinant. The modes which do not participate
in this multiplet structure arise from $\phi$ in the kernel of
$\mathscr{D}_-$ and $\Psi'$ in the kernel of $\mathscr{D}_+$.
It is easy to see from the matrix expression for ${\cal L}_\text{reg}$
that the one-loop determinant is given by the ratio of determinants of
$\mathscr{H}$ evaluated on such modes,
\begin{equation}
 Z^\text{1-loop}_\text{mat} ~=~
 \frac{\text{det}_{\Psi'}(-\mathscr{H})\big
       |_{\text{Ker}\mathscr{D}_+}}
      {\text{det}_\phi(\mathscr{H})\big
       |_{\text{Ker}\mathscr{D}_-}}.
\label{S3:eq:Zmind}
\end{equation}

The spectrum of $\mathscr{H}$ which is relevant for the above one-loop
determinant can be explicitly worked out. First, let us
consider the spectrum of $\mathscr{H}$ on the scalar $\phi$ of R-charge
$\rch$ which is annihilated by $\mathscr{D}_-$.
Assuming the form $\phi=\hat\phi(\theta)e^{im\varphi-in\chi}$, one finds
\begin{eqnarray}
 e^{i(\chi-\varphi)}\bar\epsilon\gamma^n\bar\epsilon D_n\phi &=&
 \left\{
 -\frac{b\sin\theta}{\cos\theta}(m-\rch V_\varphi)
 +\frac{\cos\theta}{b\sin\theta}(n+\rch V_\chi)
 -\frac1f\partial_\theta
 \right\}\phi~=~0,
 \nonumber \\
 \mathscr{H}\phi &=& \left\{mb+nb^{-1}-ia+\frac{Q\rch}2\right\}\phi\,.
\end{eqnarray}
The first equation determines the form of $\hat\phi(\theta)$. In
particular, from its behavior near the two ends $\theta=0$ and $\frac\pi2$,
\begin{equation}
  \hat\phi(\theta)\sim \cos^m\theta\sin^n\theta,
\end{equation}
it follows that the eigenmode is normalizable only when $m,n\ge0$. The
same analysis can be repeated for the scalar $\Psi'$ of R-charge
$(\rch-2)$ in the kernel of $\mathscr{D}_+$. We thus obtain the matter
one-loop determinant
\begin{eqnarray}
 Z^\text{1-loop}_\text{mat} &=&
 \frac{\prod_{m,n\ge0}\text{det}_R(mb+nb^{-1}+ia-\frac{Q(\rch-2)}2)}
      {\prod_{m,n\ge0}\text{det}_R(mb+nb^{-1}-ia+\frac{Q\rch}2)}
 \nonumber \\ &=&
 \prod_w s_b(\tfrac{iQ}2(1-\rch)-a\cdot w)\,,
\end{eqnarray}
where $w$ runs over all the weight vectors in the representation $R$.
This generalizes the formula (\ref{S3:eq:Zm1}) on the round sphere.

The form of the matter one-loop determinant (\ref{S3:eq:Zmind}) shows
that it can be computed from the index of the differential operators
$\mathscr{D}_\pm$
which commute with $\mathscr{H}$. In \cite{Drukker:2012sr} the relevant
index was analyzed by regarding the ellipsoid as a Hopf fibration with
the fiber direction $\partial_\varphi-\partial_\chi$. By decomposing
the fields into Fourier modes carrying different KK momentum along
the fiber, one can reduce the index to that of a differential operator
on $S^2$ and apply the fixed point formula.

Let us next consider vector multiplet. Our starting point is the
following formula for the one-loop determinant,
\begin{equation}
 Z^\text{1-loop}_\text{vec}
 ~=~
 \frac{\text{det}_\lambda(-ia-\frac1{2f}+i\,\slash\!\!\!\! D)}
      {\text{det}_{\hat A}(a^2+\Delta^{\text{vector}})^{1/2}}
 ~=~
 \frac{\text{det}_\lambda(-ia-\frac1{2f}+i\,\slash\!\!\!\! D)}
      {\text{det}_{\hat A}(-ia-\ast{\rm d})}\,,
\end{equation}
which follows from the same gauge fixing procedure as
for the round sphere (\ref{S3:eq:Zv1}). As before, the denominator is the
determinant evaluated on the space of divergenceless vector wave functions.
We evaluate this by finding out the maps between the spinor and vector
eigenmodes,
\begin{eqnarray}
&& \nu\lambda =
 -ia_\text{adj}\lambda+i\,\slash\!\!\!\! D\lambda -\frac1{2f}\lambda,
\label{S3:eq:eig1} \\
&& \nu\hat A^m =
 -ia_\text{adj}\hat A^m-\varepsilon^{mnp}\partial_n\hat A_p,\quad
 D_m\hat A^m=0.
\label{S3:eq:eig2}
\end{eqnarray}

We first notice the following identity holds for arbitrary vector field $A_m$.
\begin{equation}
 \Big(i\,\slash\!\!\!\! D-\frac1{2f}\Big)(\gamma^m\epsilon A_m)
 ~=~ i\epsilon\cdot D_mA^m
 -\gamma_m\epsilon\cdot\varepsilon^{mnp}\partial_nA_p.
\end{equation}
It follows that, for each generic vector eigenmode $\hat A_m$, one can
construct a spinor eigenmode $\lambda$ of the same eigenvalue by the map
$\lambda[\hat A]=\gamma^m\epsilon\hat A_m$. This map fails for the vector
eigenmodes satisfying $\gamma^m\epsilon\hat A_m=0$. Such modes can be
expressed in terms of a scalar $Y$ with R-charge $-2$ as,
\begin{equation}
 \hat A_m ~=~ \epsilon\gamma_m\epsilon\cdot Y.
\end{equation}
The divergence-free condition and the eigenmode equation (\ref{S3:eq:eig2})
are translated into the following conditions on $Y$,
\begin{equation}
 \mathscr{D}_+Y=0,\quad \mathscr{H}Y=\nu Y.
\end{equation}
The normalizable solutions for $Y$ are in one-to-one correspondence with
the vector eigenmodes without spinor superpartners.

Next we notice that the following identity holds for arbitrary spinor
$\lambda$,
\begin{eqnarray}
  \bar\epsilon\gamma_m\Big(
 i\,\slash\!\!\!\!D\lambda-\frac1{2f}\lambda\Big)
 {\rm d}x^m
 -i{\rm d}(\bar\epsilon\lambda)
 &=&
 -\ast{\rm d}(\bar\epsilon\gamma_m\lambda{\rm d}x^m).
\label{S3:eq:AL-id}
\end{eqnarray}
It follows that, for each generic spinor eigenmode $\lambda$, one can
construct the corresponding vector eigenmode $\hat A$ by the following map,
\begin{equation}
 \hat A[\lambda] ~=~
 (\nu+ia_\text{adj})\bar\epsilon\gamma_m\lambda {\rm d}x^m
 -i{\rm d}(\bar\epsilon\lambda)\,.
\label{S3:eq:AL}
\end{equation}
To find the kernel of this map, let us introduce two scalar functions
$\Lambda_0,\Lambda_2$ and denote
$\lambda=\epsilon\Lambda_0+\bar\epsilon\Lambda_2$. Then
$\hat A[\lambda]$ vanishes when
\begin{equation}
 \mathscr{D}_-\Lambda_0=0,\quad
 \mathscr{H}\Lambda_0=\nu\Lambda_0,\quad
 \mathscr{D}_+\Lambda_0=2(\nu+ia_\text{adj})\Lambda_2.
\label{S3:eq:kerAL}
\end{equation}
For any $\lambda$ in the kernel, one can show by applying
$\ast{\rm d}$ onto (\ref{S3:eq:AL}) that the right hand side of
(\ref{S3:eq:AL-id}) vanishes as long as $(\nu+ia_\text{adj})$ is nonzero.
Using this one can show that generic elements $\lambda$ in the kernel
automatically satisfies the eigenvalue equation (\ref{S3:eq:eig1}).
The only exceptional element in the kernel is $\lambda=\epsilon$
which does not satisfy (\ref{S3:eq:eig1}), corresponding to
$\Lambda_0=\text{const},\Lambda_2=0$ and $\nu=-ia_\text{adg}$.
The normalizable solutions to (\ref{S3:eq:kerAL}) are thus in almost
one-to-one correspondence with the spinor eigenodes without vector
superpartners.

Thus the one-loop determinant for vector multiplet can be expressed
again as the ratio of determinants of $\mathscr{H}$,
\begin{equation}
 Z^\text{1-loop}_\text{vec}~=~
 \frac{\text{det}'_{\Lambda_0}(\mathscr{H})_{\text{Ker}\mathscr{D}_-}}
      {\text{det}_{Y}(\mathscr{H})_{\text{Ker}\mathscr{D}_+}},
\end{equation}
where the prime in the enumerator indicates that the contribution from
constant modes is excluded. Apart from this minor difference, it is just
the inverse of the matter one-loop determinant for $\rch=0$, $R=\text{adj}$.
Up to an $a$-independent overall constant, we obtain
\begin{equation}
 Z^\text{1-loop}_\text{vec}
 ~=~ \prod_{\alpha\in\Delta}
  \frac{s_b(a\cdot\alpha-\tfrac{iQ}2)}{(-ia\cdot\alpha)}
 ~=~ \prod_{\alpha\in\Delta_+}
 \frac{4\sinh(\pi ba\cdot\alpha)\sinh(\pi b^{-1}a\cdot\alpha)}
      {(a\cdot\alpha)^2} \,.
\end{equation}

The general formula for the ellipsoid partition function can be
summarized as follows. Vector multiplets yield the integration measure
over Cartan subalgebra of the gauge symmetry algebra,
\begin{equation}
 \frac1{|{\cal W}|}\prod_{i=1}^r{\rm d}a_i
 \prod_{\alpha\in\Delta_+}4
 \sinh(\pi ba\cdot\alpha)
 \sinh(\pi b^{-1}a\cdot\alpha),
\end{equation}
chiral multiplets yields the determinants,
\begin{equation}
 \prod_{w\in R} s_b(\tfrac{iQ}2(1-r)-a\cdot w),
\end{equation}
and the classical Lagrangians make the following contribution to the
integrand.
\begin{equation}
 e^{-S_\text{CS}}= e^{i\pi ka\cdot a},\quad
 e^{-S_\text{FI}}=e^{4\pi i\zeta a}.
\end{equation}

Let us compare the above formula with the known result in pure
Chern-Simons theory \cite{Witten:1988hf}. Using the above formula
together with Weyl denominator formula
\begin{equation}
 \prod_{\alpha\in\Delta_+}2\sinh(\pi\alpha\cdot u)=
 \sum_{w\in{\cal W}}\epsilon(w)e^{2\pi w(\rho)\cdot u},
\end{equation}
one can express the partition function for pure SUSY Chern-Simons theory
as a sum of simple Gaussian integrals. Assuming the level $k$ to be
positive, one finds
\begin{eqnarray}
 Z_\text{CS} &=& \frac1{|\cal W|}\int\prod_{i=1}^r{\rm d}a_i
 \prod_{\alpha\in\Delta_+}
 4\sinh(\pi ba\cdot\alpha)\sinh(\pi b^{-1}a\cdot\alpha)\cdot
 \exp(i\pi k a\cdot a)
 \nonumber \\ &=&
 \exp\left(\frac{i\pi}4\text{dim}G
  +\frac{i\pi}{12k}(b^2+b^{-2})y\,\text{dim}G\right)\cdot
 k^{\frac r2}\prod_{\alpha\in\Delta_+}
 2\sin\left(\frac{\pi\alpha\cdot\rho}k\right).
\end{eqnarray}
Here $y$ is the dual Coxeter number of $G$ and $\rho$ is the Weyl vector.
We also used the formula
\begin{equation}
\rho^2=\frac1{12}\text{dim}G\,y.
\end{equation}
Apart from some phase factors, we recover the the known answer for
bosonic Chern-Simons theory at the level $k-y$. The mismatch in the
level is because there is no finite renormalization of the Chern-Simons
level for the case with ${\cal N}=2$ supersymmetry \cite{Kao:1995gf}.

\subsection{Loop observables}\label{S3:sec:loops}

Here we introduce two kinds of supersymmetric Loop operators, the Wilson
and vortex loops, and present the formulae for their expectation values.
Similar loop operators in 4D ${\cal N}=2$ theories are reviewed in
\cite{O} and play important role in understanding the AGT relation.

Supersymmetric Wilson loop operator is defined by
\begin{equation}
 W_R(C)\equiv \text{Tr}_R\text{P}\exp\oint_C
 \left(i A +\sigma {\rm d}\ell\right),
\end{equation}
where $C$ is a closed loop that winds along the direction of the Killing
vector field $\bar\epsilon\gamma^m\epsilon$, and ${\rm d}\ell$ denotes
the length element along $C$. For theories on the unit round $S^3$
where the Killing vector is along the circle fiber of Hopf fibration,
any $C$ is a great circle of radius $2\pi$. The expectation value
of Wilson loops can be calculated in the same way as partition function,
by just inserting into the integrand their classical value at the saddle
point $a$,
\begin{equation}
 W_R(C)\big|_\text{saddle}~=~ \text{Tr}_R (e^{2\pi a}).
\end{equation}
For theories on the ellipsoids with generic squashing parameter $b$
($b^2$ being irrational), the only supersymmetric closed loops are the
ones at $\theta=0$ and $\theta=\frac\pi2$ in the polar coordinate system
(\ref{S3:eq:S3bp}), since no other curves along the Killing vector
$\bar\epsilon\gamma^m\epsilon$ form closed loops. The two
choices lead to different expectation values since they have radii
$b^{-1}$ and $b$, respectively.
\begin{equation}
 W_R(\theta=0)\big|_\text{saddle}= {\rm Tr}_R(e^{2\pi a/b}),\quad
 W_R(\theta=\frac\pi2)\big|_\text{saddle} = {\rm Tr}_R (e^{2\pi ab}).
\end{equation}
There are additional supersymmetric loops for special values of the
squashing parameter. When $b=\sqrt{p/q}$ with $(p,q)$ coprime integers,
torus knots winding $p$ and $q$ times along the $\varphi$ and
$\chi$-directions at fixed $\theta\neq 0,\frac\pi2$ become supersymmetric\cite{Tanaka:2012nr}.

The vortex loop is a one-dimensional defect along which the gauge field
develops a singularity. For a vortex line lying along the $z$-axis of
the flat Euclidean $\mathbb R^3 (x,y,z)$, the gauge field strength has
delta function singularity along the line,
\begin{equation}
  F_{xy}= 2\pi H\delta(x)\delta(y)+\text{regular},
\end{equation}
where the flux $H$ takes values in the Cartan subalgebra of the gauge
symmarty algebra. 
In terms of the polar coordinate system on the $xy$-plane
$(x+iy=re^{i\theta})$, the singular behavior of the gauge field near the
vortex line is given by $A_\theta = H$. Also, it follows from
(\ref{S3:eq:STv}) that we need to impose singular boundary condition on
$D$ as well,
\begin{equation}
 D = 2\pi iH\delta(x)\delta(y)+\text{regular},
\end{equation}
in order to avoid the transformation rule of $\lambda$ and $\bar\lambda$
becoming singular.

For a vortex loop to be supersymmetric, it has to lie along the
direction of the Killing vector $\bar\epsilon\gamma^m\epsilon$.
We orient the vortex loops so that the $+z$ direction always agrees with
the direction of Killing vector and define the flux $H$ accordingly.
For generic ellipsoid backgrounds, supersymmetric vortex loops can only
lie along the direction of $(-\varphi)$ at $\theta=0$, or the direction
of $(+\chi)$ at $\theta=\frac\pi2$. These two vortex loops are expressed
by the flat gauge fields,
\begin{equation}
 (\theta=0)~~ A=H{\rm d}\chi,\qquad
 (\theta=\frac\pi2)~~ A=-H{\rm d}\varphi.
\end{equation}

Let us hereafter restrict the discussion to the vortex loops in abelian
gauge theory and evaluate their expectation value. First, notice that the
introduction of a vortex loop with flux $H$ in Chern-Simons theory at
level $k$ induces a Wilson loop with charge $-kH$. To see this, let us
decompose the vector multipet fields in the presence of a vortex loop
into the singular and regular parts,
$\mathscr{A}=\mathscr{A}_\text{sing}+\mathscr{A}_\text{reg}$. Then the SUSY
Chern-Simons action integral for such $\mathscr{A}$ becomes
\begin{equation}
 S_\text{CS}[\mathscr{A}_\text{sing}+\mathscr{A}_\text{reg}]~=~
 ikH\oint_C (A_\text{reg}-i\sigma_\text{reg}{\rm d}\ell)
 + S_\text{CS}[\mathscr{A}_\text{reg}]\,.
\end{equation}
Therefore, the value of classical Chern-Simons action at the saddle
point $a$ gets shifted because of the vortex loop as
\begin{equation}
 e^{i\pi k (a^2+2iab^{-1}H)}\quad\text{or}\quad
 e^{i\pi k (a^2+2iabH)}.
\end{equation}
The value of the FI term $e^{-S_\text{FI}}=e^{4\pi i\zeta a}$ remains the
same. Now one can go through the evaluation of the one-loop determinant
again, where the only difference is that there is a nonzero flat gauge
field in addition to a constant scalar $a$. Since it enters in the operator
$\mathscr{H}$ as follows,
\begin{equation}
 \mathscr{H}\phi ~=~
 \left\{-ib(\partial_\varphi-iA_\varphi)
 +ib^{-1}(\partial_\chi-iA_\chi)-ia+\frac{Qr}2\right\}\phi,
\end{equation}
the effect of the vortex loop can be incorporated by shifting $a$ in
our previous formula by $-ibA_\varphi+ib^{-1}A_\chi$. Depending on
whether the vortex loop is put at $\theta=0$ or $\frac\pi2$, the saddle
point parameter $a$ is shifted by $ib^{-1}H$ or $ibH$.

Since the parameter $a$ is to be integrated over, the shift of $a$ by
$ib^{\pm1}H$ can be undone by shifting its integration contour. This
also eliminates the shift of classical Chern-Simons action by a Wilson
line. As a result, the effect of a vortex loop of flux
$H$ in abelian Chern-Simons theory at level $k$, FI coupling $\zeta$
just amounts to a multiplication of the factor
\begin{equation}
 \exp\left(i\pi k b^{\mp2}H^2 + 4\pi\zeta b^{\mp1}H\right)\,.
\end{equation}

Our argument so far assumed that $H$ is small. The computation
of one-loop determinants on ellipsoids was based on the spectrum of
normalizable eigenmodes of $\mathscr{H}$ in the kernel of the operators
$\mathscr{D}_\pm$, but normalizability of the eigenmodes is affected
by nonzero $H$. Also, the shift of $a$-integration contour may hit poles
in the integrand. See \cite{Kapustin:2012iw,Drukker:2012sr} for further
discussions.

Vortex loops can also be introduced for flavor symmetry of matter
chiral multiplets, by coupling the corresponding current to a singular
background gauge field with nonzero flux $H$ localized along a loop. Its
effect is similar to that of real mass deformation, namely we have the
appearance of $ib^{\mp1}H$ in place of the real mass $m$ in the matter
one-loop determinants.

\section{4D Superconformal Index}\label{S3:sec:4DInd}

Superconformal index was introduced for 4D ${\cal N}=1$ superconformal
field theories by R\"omelsberger \cite{Romelsberger:2005eg,Romelsberger:2007ec}
and for more general cases by Kinney et.al. \cite{Kinney:2005ej},
as a quantity which encodes the spectrum of BPS operators.
In superconformal theories, the spectrum of BPS operators is in
correspondence with the spectrum of states in radial quantization.
The index can therefore be formulated in terms of path integral on
$S^1\times S^3$, with an appropriate periodicity condition along the
$S^1$. The periodicity can be twisted by various symmetries of the
theory in such a way to preserve part of SUSY. The index is then a
function of the fugacity variables that parametrize the twist.

The superconformal index is invariant under any SUSY-preserving
continuous deformation of the theory and, in particular, independent
of the gauge coupling. The indices of nontrivial theories at the RG
fixed point can therefore be evaluated using the weak coupling
description at high energy where saddle point approximation becomes exact.

Here we present the path integral derivation of the superconformal index
for 4D ${\cal N}=1$ SUSY theories. Our purpose here is to explain the
connection between 3D partition functions on $S^3$ and 4D superconformal
indices which was studied in \cite{Dolan:2011rp,Gadde:2011ia,Imamura:2011uw}.
Interestingly, some of the fugacity variables turn into parameters of
supersymmetric deformations of the round $S^3$ upon dimensional reduction.
As an important example, we reproduce two inequivalent SUSY backgrounds
which are both based on the same squashed $S^3$ with $SU(2)\times U(1)$
isometry but characterized by different Killing spinor equations
\cite{Hama:2011ea,Imamura:2011wg}.

The superconformal indices for 4D ${\cal N}=2$ theories of class S are
in correspondence with partition function of 2D $q$-deformed Yang-Mills
theory, as reviewed in \cite{RR} in this volume.

\subsection{4D ${\cal N}=1$ SUSY theories}\label{S3:sec:4Dtheory}

We again begin by fixing the notations. In four dimensions there
are two kinds of doublet spinors $\psi_\alpha$ and $\bar\psi^{\dot\alpha}$,
corresponding to two copies of $SU(2)$ that form the 4D rotation
symmetry. Their spinor indices are raised or lowered by antisymmetric
$\epsilon$ tensors with nonzero elements $\epsilon^{12}=-\epsilon_{12}=1$.
We introduce the $2\times2$ matrices,
\begin{equation}
 \sigma_a=\bar\sigma_a=\text{Pauli matrix}~(a=1,2,3);\quad
 \sigma_4=i,\quad\bar\sigma_4=-i,
\end{equation}
with index structure $(\sigma_a)_{\alpha\dot\beta}$ and
$(\bar\sigma_a)^{\dot\alpha\beta}$, satisfying standard algebra.
We also use
$\sigma_{ab}\equiv\frac12(\sigma_a\bar\sigma_b-\sigma_b\bar\sigma_a)$ and
$\bar\sigma_{ab}\equiv\frac12(\bar\sigma_a\sigma_b-\bar\sigma_b\sigma_a)$.

Although 4D ${\cal N}=1$ supersymmetric theories on general curved
backgrounds and the equations for Killing spinors can be obtained from
off-shell supergravity \cite{Festuccia:2011ws}, here we take a heuristic
approach. We consider the following Killing spinor equation,
\begin{equation}
 D_m\epsilon = \sigma_m\bar\kappa,\quad
 D_m\bar\epsilon = \bar\sigma_m\kappa\quad
\text{for some }\kappa,\bar\kappa.
\end{equation}
where the covariant derivative $D_m$ contains the gauge field $V_m$ for
$U(1)_\text{R}$ under which $\epsilon,\bar\epsilon$ are charged $+1,-1$.
Using these Killing spinors we set the transformation
rule for ${\cal N}=1$ vector multiplets,
\begin{eqnarray}
 \delta A_m &=&
 \frac i2(\epsilon\sigma_m\bar\lambda-\bar\epsilon\bar\sigma_m\lambda),
 \nonumber \\
 \delta\lambda &=& \frac12\sigma^{mn}\epsilon F_{mn}-\epsilon D,
 \nonumber \\
 \delta\bar\lambda &=& \frac12\bar\sigma^{mn}\bar\epsilon F_{mn}
 +\bar\epsilon D,
 \nonumber \\
 \delta D &=&
 -\frac i2\epsilon\sigma^mD_m\bar\lambda
 -\frac i2\bar\epsilon\bar\sigma^mD_m\lambda,
\end{eqnarray}
and chiral multiplets,
\begin{eqnarray}
 \delta\phi &=& -\epsilon\psi,\quad
 \delta\psi ~=~ i\sigma^m\bar\epsilon D_m\phi
 +\frac{3i\rch}4\sigma^mD_m\bar\epsilon\phi +\epsilon F,
 \nonumber \\
 \delta\bar\phi &=& +\bar\epsilon\bar\psi,\quad
 \delta\bar\psi ~=~ i\bar\sigma^m\epsilon D_m\bar\phi
 +\frac{3i\rch}4\bar\sigma^mD_m\epsilon\bar\phi +\bar\epsilon\bar F,
 \nonumber \\
 \delta F &=& i\bar\epsilon\bar\sigma^mD_m\psi
 +\frac{i(3\rch-2)}4D_m\bar\epsilon\bar\sigma^m\psi
 -i\bar\epsilon\bar\lambda\phi,
 \nonumber \\
 \delta\bar F &=& -i\epsilon\sigma^mD_m\bar\psi
 -\frac{i(3\rch-2)}4D_m\epsilon\sigma^m\bar\psi
 -i\epsilon\bar\phi\lambda.
\end{eqnarray}
Here $\rch$ is the R-charge of the field $\phi$. The scaling weight and the
R-charge of the fields are summarized in the table \ref{S3:tab:wr4}.
\begin{table}
\def\STRUT{\rule[-0.85mm]{0mm}{5.5mm}}
\begin{center}
\begin{tabular}{c||cc|cccc|cccccc}
\hline
fields & $\epsilon$ & $\bar\epsilon$ &
 $A_a$ & $\lambda$ & $\bar\lambda$ & $D$ &
 $\phi$ & $\bar\phi$ & $\psi$ & $\bar\psi$ & $F$ & $\bar F$ \STRUT\\
\hline
weight & $-\frac12$ & $-\frac12$ &
 $1$ & $\frac32$ & $\frac32$ & $2$ &
 $\frac{3\rch}2$ & $\frac{3\rch}2$ & $\frac{3\rch+1}2$ & $\frac{3\rch+1}2$ &
 $\frac{3\rch+2}2$ & $\frac{3\rch+2}2$ \STRUT\\
\hline
R-charge & $1$ & $-1$ &
 $0$ & $1$ & $-1$ & $0$ &
 $\rch$ & $-\rch$ & $\rch-1$ & $1-\rch$ & $\rch-2$ & $2-\rch$ \STRUT\\
\hline
\end{tabular}
\caption{the scaling weight and the R-charge of the fields.}
\label{S3:tab:wr4}
\end{center}
\end{table}

Given a 3D background ${\cal M}$ with a pair of Killing spinors
$\epsilon,\bar\epsilon$ satisfying (\ref{S3:eq:KS1}) and
(\ref{S3:eq:KS1-2}), one can construct a 4D ${\cal N}=1$ supersymmetric
background ${\cal M}\times \mathbb R$ by choosing the metric and the
$U(1)_\text{R}$ gauge field as follows.
\begin{equation}
 {\rm d}s^2_\text{(4D)}= e^ae^a={\rm d}s^2_{{\cal M}} + {\rm d}t^2
 ~~(e^4\equiv {\rm d}t),\quad
 V_\text{(4D)} = V_\text{(3D)} -iM{\rm d}t.
\end{equation}
The 3D Killing spinors $\epsilon,\bar\epsilon$ are promoted to 4D
Killing spinors satisfying
\begin{equation}
 D_m\epsilon = -M\sigma_m\bar\sigma_4\epsilon,\quad
 D_m\bar\epsilon = M\bar\sigma_m\sigma_4\bar\epsilon.
\label{S3:eq:KS4}
\end{equation}
The following supersymmetric Lagrangians on this background are
relevant in the computation of the index.
\begin{eqnarray}
 {\cal L}_\text{g} &=&
 {\rm Tr}\Big(\frac12F_{mn}F^{mn}
 +D^2+i\bar\lambda\bar\sigma^mD_m\lambda\Big),
 \nonumber \\
 {\cal L}_\text{m} &=&
 D_m\bar\phi D^m\phi+(3\rch-2)M(D_4\bar\phi\phi-\bar\phi D_4\phi)
 +\Big\{\frac{\rch R}4-3\rch(3\rch-2)M^2\Big\}\bar\phi\phi -i\bar\phi D\phi
 \nonumber \\ &&
 -i\bar\psi\bar\sigma^mD_m\psi -i(3\rch-2)M\bar\psi\bar\sigma_4\psi
 +i\bar\psi\bar\lambda\phi+i\bar\phi\lambda\psi
 +\bar FF.
\label{S3:eq:L4}
\end{eqnarray}
Here $D_4$ is the fourth component of $D_a\equiv e_a^mD_m$.

It is a useful observation that the above 4D transformation rules and
Lagrangians can actually be obtained from the corresponding 3D
quantities by the simple replacement
$\sigma\to A_t+i\partial_t$.

\subsection{Path integral formulation of the index}\label{S3:sec:PIInd}

Let us choose ${\cal M}$ to be the unit round sphere and set
$M=\frac12, V=-\frac i2{\rm d}t$. The Killing spinor equation (\ref{S3:eq:KS4}) on
this background has two independent solutions for each of $\epsilon$ and
$\bar\epsilon$, which are all constant spinors in the left-invariant
frame. Besides these four solutions,
there are four solutions to (\ref{S3:eq:KS4}) with the right hand side
sign-flipped. These eight solutions correspond to the eight
supercharges in the 4D ${\cal N}=1$ superconformal algebra, but the
Lagrangians in (\ref{S3:eq:L4}) with $M=\frac12$ are invariant only under the
first four.

From the four Killing spinors satisfying (\ref{S3:eq:KS4}), let us pick up
the two characterized by $\gamma_3\epsilon=-\epsilon$ and
$\gamma_3\bar\epsilon=\bar\epsilon$, and denote the corresponding
supercharges by $\bf S$ and $\bf Q$. The R-charges and $SU(2)_\mathscr{R}$
spins of ${\bf S,Q}$ are opposite to those of the corresponding Killing
spinors, so $\bf S$ has ${\bf R}=-1, {\bf J}^3_\mathscr{R}=+\frac12$
while $\bf Q$ has ${\bf R}=1, {\bf J}^3_\mathscr{R}=-\frac12$. The
anticommutator of $\bf S$ and $\bf Q$ can be found from the algebra of the
corresponding SUSY transformations acting on fields. With a suitable
normalization of $\epsilon,\bar\epsilon$ one finds
\begin{equation}
 \{{\bf S},{\bf Q}\} = -\partial_t +iA_t-2{\bf J}^3_\mathscr{R}-{\bf R},
\end{equation}
where $A_t$ is the component of dynamical gauge field and
${\bf J}^a_\mathscr{R}$ is the sum of isometry rotation of $S^3$ and
local Lorentz rotation. Note that, since we have turned on the
background $U(1)_\text{R}$ gauge field so that the Killing spinors
corresponding to $\bf S,Q$ are time independent, the time derivative
$-\partial_t+iA_t$ should not be simply related to the dilation $\bf D$.
Rather it should be identified with ${\bf D}-\frac12{\bf R}$ which
commutes with the supercharges $\bf S$ and $\bf Q$. Thus we have
reproduced an important subalgebra of the 4D ${\cal N}=1$ superconformal
algebra,
\begin{equation}
 \{{\bf S},{\bf Q}\} = {\bf D}-2{\bf J}^3_\mathscr{R}-\frac32{\bf R}
 \equiv {\bf H} .
\end{equation}

Now let us compactify the time direction $t\sim t+\beta$. The path
integral on the resulting background $S^3\times S^1$ defines the
superconformal index. In the simplest example where all the fields obey
periodic boundary condition, one obtains
\begin{equation}
 I ~=~ {\rm Tr}[(-1)^{\bf F}q^{{\bf D}-\frac12{\bf R}}].\quad
 (q\equiv e^{-\beta})
\end{equation}
This form can be generalized by twisting the periodicity of fields by
various symmetries which commute with the supercharges $\bf S,Q$.
Some of such symmetries are in the superconformal algebra. The Cartan
subalgebra of its bosonic part is generated by the dilation $\bf D$, the
$U(1)$ R-charge $\bf R$ and the two rotation generators
${\bf J}^3_\mathscr{L}, {\bf J}^3_\mathscr{R}$, of which three linear
combinations commute with $\bf S$ and $\bf Q$.
Also, in theories with additional global symmetry, one can use any of its
elements $\bf m$ to modify the periodicity.
The fully generalized index is then given by
\begin{equation}
 I ~=~ {\rm Tr}[(-1)^{\bf F}
 q^{{\bf D}-\frac12{\bf R}}x^{2{\bf J}^3_\mathscr{R}+{\bf R}}
 y^{2{\bf J}^3_\mathscr{L}}e^{i{\bf m}\beta}],\quad
 q=e^{-\beta}, x=e^{i\beta\xi}, y=e^{i\beta\eta}
\label{S3:eq:Ind}
\end{equation}
and is a function of the fugacity parameters $\xi,\eta$ and $\bf m$ as
well as $\beta$. An important remark here is that the only states which
contribute to the index are those annihilated by the supercharges
$\bf Q, \bf S$ and also by their anticommutator $\bf H$. The
index therefore depends on $q$ and $x$ only through their product
$qx=e^{-\beta+i\beta\xi}$.

The index (\ref{S3:eq:Ind}) is given by a path integral over fields obeying
twisted periodicity condition. By a suitable field redefinition, it can be
rewritten into a path integral over ordinary periodic fields but with a
deformed Lagrangian. In this process, the twists by R- or flavor
symmetries turn into a constant background gauge fields along the $t$
direction. On the other hand, the twist by rotational symmetries means
Scherk-Schwarz like compactification,
\begin{equation}
 (t,g)~\sim~ (t+\beta, e^{-i\beta\eta\gamma^3}ge^{i\beta\xi\gamma^3}).
\end{equation}
This can be brought into a system with ordinary time periodicity by a
suitable change of coordinates, but then the metric written in the new
coordinates aquires off-diagonal components
\begin{equation}
 {\rm d}s^2 ~=~ {\rm d}t^2 +
 g^{(S^3)}_{mn}({\rm d}x^m+u^m{\rm d}t)({\rm d}x^n+u^n{\rm d}t),
 \quad
 u\equiv 2i\xi\mathscr{R}^3+2i\eta\mathscr{L}^3.
\end{equation}
Here the vector fields $\mathscr{L}^a, \mathscr{R}^a$ are properly
normalized generators of $SU(2)_\mathscr{L,R}$. In fact,
the effect of this deformation of the metric on field theory is simply to
modify the time derivative $\partial_t$ by the rotation generator. For
example, the kinetic term for a free scalar becomes
\begin{equation}
  \frac12(\partial_t\phi-u^m\partial_m\phi)^2
 +\frac12g^{mn}_{(S^3)}\partial_m\phi\partial_n\phi.
\end{equation}
A little more work shows that, for spinor fields, the time derivative is
modified by a combination of $u^m\partial_m$ and a local Lorentz
transformation which makes precisely the action of the rotation symmetry.
Summarizing, the general index (\ref{S3:eq:Ind}) can be computed by path integral
over periodic fields on $S^1\times S^3$, with the following replacement
in the Lagrangian (\ref{S3:eq:L4})
\begin{equation}
 i\partial_t ~\longmapsto~
 i\hat\partial_t\equiv i\partial_t+\xi(2{\bf J}^3_\mathscr{R}+{\bf R})
 +2\eta {\bf J}^3_\mathscr{L}+{\bf m}.
\label{S3:eq:dt}
\end{equation}

\subsection{Evaluation of the index}\label{S3:sec:EVInd}

Let us turn to the evaluation of the index. Since the index is
invariant under deformations preserving the algebra of
${\bf S}, {\bf Q}, {\bf H}$, we introduce the sum of ${\cal L}_\text{g}$
and ${\cal L}_\text{m}$ in (\ref{S3:eq:L4}) with a large overall
coefficient into the path integral weight so that the argument of
exact saddle point analysis apply. This time, the saddle points are
labeled by the constant value of gauge field along time direction $A_t=a$.

Let us evaluate the one-loop determinant, first for the
vector multiplet with gauge group $G$. It is most convenient to work
in the temporal gauge $A_t=a$, for which we need to introduce ghosts
with kinetic term ${\rm Tr}(\bar cD_tc)$. The gaussian integral over
fluctuations gives
\begin{equation}
\frac
{\text{Det}_\lambda (\hat\partial_t-ia-\frac12+i\;\slash\!\!\!\!D_{S^3})
 \text{Det}'_c(\hat\partial_t-ia)}
{\text{Det}_A (-(\hat\partial_t-ia)^2+\Delta_{S^3}^\text{vector})^{\frac12}}\,.
\end{equation}
Here the prime indicates that the constant modes of the ghosts are
excluded, and $\hat\partial_t$ is defined in (\ref{S3:eq:dt}). Expanding
the fields into
spherical harmonics which diagonalizes the Laplace or Dirac operators on
$S^3$, the above determinant can be rewritten into an infinite product
of 1D Dirac determinants on the circle of circumference $\beta$,
\begin{equation}
 \text{det}(\partial_t-ix)~=~
 \prod_{k\in\mathbb Z}(2\pi ik\beta^{-1}-ix)~=~
 -2i\sin\frac{\beta x}2.
\end{equation}
The integral over the ghost modes with
$SU(2)_\mathscr{L}\times SU(2)_\mathscr{R}$ spin $(0,0)$ yields
\begin{equation}
 \text{det}'_\text{adj}(\partial_t-ia)
~=~ \beta^{\text{rk}G}\prod_{\alpha\in\Delta_+}
   \Big(\frac{2\sin(\beta\alpha\cdot a/2)}{\alpha\cdot a}\Big)^2,
\end{equation}
where we assumed $a$ to take values in Cartan torus.
Combined with the Vandermonde determinant, this gives an appropriate
measure factor for the integration over Cartan torus.
\begin{equation}
 {\rm d}\mu(a)~=~
 \frac1{|{\cal W}|}\prod_{i=1}^r\frac{{\rm d}\hat a_i}{2\pi}
 \prod_{\alpha\in\Delta_+}4\sin^2\frac{\alpha\cdot\hat a}2.\quad
 (\hat a\equiv\beta a)
\label{S3:eq:mua}
\end{equation}
The integral over the remaining modes of all the fields gives, after an
enormous cancellation between bosonic and fermionic contributions,
the following.
\begin{eqnarray}
 I_\text{vec} &=&
 \prod_{n\ge1}
 \text{det}_\text{adj}\big(\partial_t-ia+n(1-i\xi+i\eta)\big)
 \text{det}_\text{adj}\big(\partial_t-ia+n(1-i\xi-i\eta)\big).
 \nonumber \\
 &=&
 I_0(q_1,q_2)^{\text{rk}G}\cdot
 \prod_{\alpha\in\Delta,n\ge1}
 (1-q_1^ne^{i\alpha\cdot\hat a})
 (1-q_2^ne^{i\alpha\cdot\hat a}),
\label{S3:eq:Iv}
\end{eqnarray}
where
\begin{eqnarray}
 && I_0(q_1,q_2) ~\equiv~ 
\prod_{n\ge1}(1-q_1^n)(1-q_2^n),
 \nonumber \\
 &&
 q_1\equiv qxy = e^{-\beta(1-i\xi-i\eta)},\quad
 q_2\equiv qx/y = e^{-\beta(1-i\xi+i\eta)}.
\end{eqnarray}
The first line in (\ref{S3:eq:Iv}) can be regarded as a refinement of
the 3D result (\ref{S3:eq:Zv2}) corresponding to the addition of one more
dimension with periodicity $\beta$ and twists $\xi,\eta$.
Note that, in going to the second line, an infinite zero-point energy
has been regularized so that the result agree with what we would obtain
from canonical quantization.

To compute the index from canonical formalism, we decompose the
vector multiplet fields on $S^1\times S^3$ using spherical harmonics
and reduce the free super-Yang-Mills theory to a quantum mechanics
of infinitely many bosonic and fermionic harmonic oscillators.
The oscillator modes all carry definite eigenvalues of
${\bf R}, {\bf J}_\mathscr{L}^3, {\bf J}_\mathscr{R}^3$, and their
frequency determines the eigenvalue of ${\bf D}-\frac12{\bf R}$.
In computing the index as a trace over the Fock space, it is convenient
to first consider the trace over one-particle states called the letter
index. For a vector multiplet for gauge group $G$ it is given by
\begin{eqnarray}
 i_\text{vec} &\equiv& \text{Tr}_\text{(1p)}\left[
  (-1)^{\bf F}q^{{\bf D}-\frac12{\bf R}}
  x^{2{\bf J}^3_\mathscr{R}+{\bf R}}
  y^{2{\bf J}^3_\mathscr{L}}
  e^{i\hat a}
 \right]
 \nonumber \\ &=&
 \text{tr}_\text{adj}U\cdot
 \sum_{n\ge0}\left(
 q^{n+2}\chi_{\frac{n+2}2,\frac n2}
+q^{n+2}\chi_{\frac n2,\frac{n+2}2}
-xq^{n+1}\chi_{\frac{n+1}2,\frac n2}
-x^{-1}q^{n+2}\chi_{\frac n2,\frac{n+1}2}\right),
 \nonumber \\ &&
U\equiv e^{i\hat a}\in G,\quad
\chi_{j,\bar j}\equiv\text{tr}_{(j,\bar j)}
[x^{2{\bf J}^3_\mathscr{R}}y^{2{\bf J}^3_\mathscr{L}}].
\end{eqnarray}
In fact, all the oscillators not saturating the bound ${\bf H}\ge0$
form pairs and do not contribute to the letter index.
Indeed, the above letter index can be simplified as follows,
\begin{equation}
 i_\text{vec}
 ~=~ -\left(\frac{q_1}{1-q_1}+\frac{q_2}{1-q_2}\right)\text{tr}_\text{adj}(U)
 \,.
\end{equation}
The full index is then obtained as its plethystic exponential,
\begin{equation}
 I_\text{vec} ~=~
 \text{PE}\Big[
 i_\text{vec}(q_1,q_2,U)\Big]
 ~\equiv~
 \exp\left(\sum_{n\ge1}\frac 1n\, i_\text{vec}(q_1^n,q_2^n,U^n)\right),
\end{equation}
integrated over $U$ in the Cartan torus with the invariant measure
(\ref{S3:eq:mua}).

Let us next consider the chiral multiplet of R-charge $\rch$ in the
representation $R$ of the gauge group. Its one-loop determinant is
\begin{eqnarray}
I_\text{mat} &=&
\frac
 {\text{Det}_\psi(\hat\partial_t-ia+\rch-\frac12+i\;\slash\!\!\!\! D_{S^3})}
 {\text{Det}_\phi(-(\hat\partial_t-ia+\rch-1)^2+\Delta^\text{scalar}_{S^3}+1)}
\nonumber \\ &=&
\prod_{m,n\ge0}\frac
{\text{det}_R(-\partial_t+ia+(1-i\xi)(m+n+2-\rch)-i\eta(m-n))}
{\text{det}_R(\partial_t-ia+(1-i\xi)(m+n+\rch)-i\eta(m-n))}\,.
\end{eqnarray}
This can again be regarded as a refinement of the one-loop determinant
(\ref{S3:eq:Zm1}) for 3D chiral multiplet. With an appropriate
regularization of the zero-point energy, one can rewrite this further
as a product over the weights of the representation $R$,
\begin{equation}
 I_\text{mat} ~=~ \prod_w \Gamma(e^{iw\cdot\hat a}
 (q_1q_2)^{\frac \rch2};q_1,q_2),
\end{equation}
where $\Gamma(z;q_1,q_2)$ is the elliptic Gamma function
\begin{equation}
 \Gamma(z;q_1,q_2)~=~ \prod_{m,n\ge0}
 \frac{1-z^{-1}q_1^{m+1}q_2^{n+1}}{1-zq_1^mq_2^n}\,.
\end{equation}
This result can also be obtained from canonical formalism, as the
plethystic exponential of the letter index,
\begin{eqnarray}
 i_\text{mat} &=&
 \text{tr}_R(U)\cdot
 \sum_{n\ge0}\left(
 x^rq^{n+r}\chi_{\frac n2,\frac n2}
-x^{r-1}q^{n+r+1}\chi_{\frac {n+1}2,\frac n2}
 \right)
 \nonumber \\ && +
 \text{tr}_{\bar R}(U)\cdot
 \sum_{n\ge0}\left(
 x^{-r}q^{n+2-r}\chi_{\frac n2,\frac n2}
-x^{1-r}q^{n+r+1}\chi_{\frac n2,\frac{n+1}2}
 \right)
 \nonumber \\ &=&
 \frac{(q_1q_2)^{\frac\rch2}\text{tr}_R(U)
      -(q_1q_2)^{1-\frac\rch2}\text{tr}_R(U^{-1})}
      {(1-q_1)(1-q_2)}.
\end{eqnarray}

\subsection{Squashed $S^3$ from twisted compactifications}\label{S3:sec:SScpt}

In the limit $\beta\to0$ where one can neglect the KK modes, the 4D
superconformal index reduces to 3D partition function, but with a new
dependence on additional parameters $\xi,\eta$. They enter into the 3D
partition function through the squashing parameter $b$,
\begin{equation}
 b^2=\frac{1-i\xi+i\eta}{1-i\xi-i\eta}.
\end{equation}
Recall that we have chosen the background $S^3\times S^1$ with
$M=\frac12$ at the beginning of Section \ref{S3:sec:PIInd}, and that our
computation was preserving a pair of
left-invariant supercharges $\bf Q,S$. In this case, the above relation
shows that the twist by ${\bf J}^3_\mathscr{R}$ (accompanied by an appropriate
R-twist) has a rather trivial effect on the partition function, but the
twist by ${\bf J}^3_\mathscr{L}$ does change the partition function
in a non-trivial manner. So the different Scherk-Schwarz twists lead to
qualitatively different 3D backgrounds after dimensional reduction.

To understand the effect of two different twists upon 3D geometry, let
us consider instead the twisted compactification with $\xi\ne0, \eta=0$ and
try different choices of unbroken supersymmetry. After moving to the
coordinate system with ordinary time periodicity, the metric is given by
\begin{equation}
  {\rm d}s^2 ~=~ E^aE^a ~=~ e^1e^1+e^2e^2+(e^3+\xi {\rm d}t)^2 + {\rm d}t^2.
\label{S3:eq:sscpt}
\end{equation}
On this space, one can either preserve left-invariant or right-invariant
supercharges by choosing the background $U(1)_\text{R}$ gauge field
appropriately to make the corresponding Killing spinors $t$-independent.
For $V_t=-\frac i2+\xi$, the Killing spinor equation (\ref{S3:eq:KS4}) with
$M=\frac12$ has a pair of time-independent solutions satisfying
$\gamma_3\epsilon=-\epsilon$ and $\gamma_3\bar\epsilon=+\bar\epsilon$,
which we identified with the left-invariant supercharges ${\bf S},{\bf Q}$.
The solutions corresponding to the other pair of left-invariant
supercharges become time-independent when $V_t=-\frac i2-\xi$. For
$V_t=\frac i2$, the Killing spinor equation (\ref{S3:eq:KS4}) with $M=-\frac12$
has solutions corresponding to the four right-invariant supercharges.

To do the dimensional reduction along $S^1$, we rewrite the metric
(\ref{S3:eq:sscpt}) into the form
\begin{equation}
  {\rm d}s^2 ~=~ \hat E^a\hat E^a~=~
 e^1e^1+e^2e^2+u^2e^3e^3
 +u^{-2}({\rm dt}+u^2\xi\, e^3)^2,
\label{S3:eq:sqS3}
\end{equation}
where $u\equiv (1+\xi^2)^{-1/2}$. Since this can be regarded as a local
Lorentz transformation, the Killing spinors on the new local Lorentz
frame satisfy
\begin{equation}
D_m\epsilon=-M\sigma_m(\bar\sigma_ah^a)\epsilon,\quad
D_m\bar\epsilon=M\bar\sigma_m(\sigma_ah^a)
 \bar\epsilon,
\end{equation}
where $M=\frac12$ or $-\frac12$ for the left- or right-invariant Killing
spinors, and
\begin{equation}
 h^a=(0,0,-u\xi,u).
\end{equation}
By dropping the last term on the right hand side of (\ref{S3:eq:sqS3}) we
obtain the 3D metric of the familiar squashed $S^3$ with
$SU(2)_\mathscr{L}\times U(1)_\mathscr{R}$ isometry.
But the nature of the dimensionally reduced theory depends also on which
supersymmetries have been preserved in the reduction.

If we set $M=\frac12$ and $V_t=-\frac i2+\xi$ upon dimensional reduction, the
supersymmetry of the resulting 3D theory is characterized by the Killing
spinor equation
\begin{eqnarray}
 \Big(\partial_m+\frac14\omega_m^{ab}\gamma^{ab}+iu\xi^2 V_m\Big)\epsilon
 &=& \frac{iu}2\gamma_m\epsilon,
 \nonumber \\
 \Big(\partial_m+\frac14\omega_m^{ab}\gamma^{ab}-iu\xi^2 V_m\Big)\bar\epsilon
 &=& \frac{iu}2\gamma_m\bar\epsilon,
\end{eqnarray}
where $V_m\equiv \hat E^3_m= u e^3_m$. The above Killing spinor equation
takes the form of (\ref{S3:eq:KS1}) with $U_m=0$, and $1/4$ of the supersymmetry
on the round $S^3$ remains unbroken after squashing due to the background
$U(1)_\text{R}$ gauge field $-u\xi^2 V_m$. It was shown in
\cite{Hama:2011ea} that the exact partition function on this squashed
$S^3$ background is essentially the same as that on the round $S^3$, in
consistency with the discussion in the previous subsection. For the case
$M=\frac12$ and $V_t=-\frac i2+\xi$, the 3D Killing spinor equation takes the
same form as above but the $U(1)_\text{R}$ gauge field appears with the
opposite sign.

If we set $M=-\frac12$ and $V_t=\frac i2$, the Killing spinor equation
of the 3D theory is
\begin{eqnarray}
 \Big(\partial_m+\frac14\omega_m^{ab}\gamma^{ab}\Big)\epsilon
 &=& -\frac{iu}2\gamma_m\epsilon-u\xi V^n\gamma_{mn}\epsilon,
 \nonumber \\
 \Big(\partial_m+\frac14\omega_m^{ab}\gamma^{ab}\Big)\bar\epsilon
 &=& -\frac{iu}2\gamma_m\bar\epsilon+u\xi V^n\gamma_{mn}\bar\epsilon,
\end{eqnarray}
again with $V_m\equiv \hat E^3_m= u e^3_m$. This case preserves 1/2 of
the Killing spinors on the round $S^3$. The above Killing spinor
equation can be identified with (\ref{S3:eq:KS1}) with $U_m\ne 0$.
It was shown in \cite{Imamura:2011wg} that the partition function on
this background depends nontrivially on $\xi$ through the squashing parameter
\begin{equation}
 b=u(1-i\xi).
\end{equation}
For a real $\xi$, the squashing parameter $b$ is a complex phase.

\subsection*{Acknowledgments}

The author thanks Naofumi Hama, Sungjay Lee and Jaemo Park for
collaboration on the materials discussed in this article. The author
also thanks the string theory group at Yukawa Institute for Theoretical
Physics, Kyoto University where the major part of this article was written.

\bibliographystyle{utphys}
\small\baselineskip=.93\baselineskip
\let\bbb\bibitem\def\bibitem{\itemsep1pt\bbb}
\bibliography{bib}

\providecommand{\href}[2]{#2}\begingroup\raggedright\begin{thebibliography}{10}

\bibitem{Kapustin:2009kz}
A.~Kapustin, B.~Willett, and I.~Yaakov, ``{Exact Results for Wilson Loops in
  Superconformal Chern-Simons Theories with Matter},''
  \href{http://dx.doi.org/10.1007/JHEP03(2010)089}{{\em JHEP} {\bfseries 1003}
  (2010) 089},
\href{http://arxiv.org/abs/0909.4559}{{\ttfamily arXiv:0909.4559 [hep-th]}}.

\bibitem{Jafferis:2010un}
D.~L. Jafferis, ``{The Exact Superconformal R-Symmetry Extremizes Z},''
  \href{http://dx.doi.org/10.1007/JHEP05(2012)159}{{\em JHEP} {\bfseries 1205}
  (2012) 159},
\href{http://arxiv.org/abs/1012.3210}{{\ttfamily arXiv:1012.3210 [hep-th]}}.

\bibitem{Hama:2010av}
N.~Hama, K.~Hosomichi, and S.~Lee, ``{Notes on SUSY Gauge Theories on
  Three-Sphere},'' \href{http://dx.doi.org/10.1007/JHEP03(2011)127}{{\em JHEP}
  {\bfseries 1103} (2011) 127},
\href{http://arxiv.org/abs/1012.3512}{{\ttfamily arXiv:1012.3512 [hep-th]}}.

\bibitem{Pestun:2007rz}
V.~Pestun, ``{Localization of gauge theory on a four-sphere and supersymmetric
  Wilson loops},'' \href{http://dx.doi.org/10.1007/s00220-012-1485-0}{{\em
  Commun.Math.Phys.} {\bfseries 313} (2012) 71--129},
\href{http://arxiv.org/abs/0712.2824}{{\ttfamily arXiv:0712.2824 [hep-th]}}.

\bibitem{Hama:2011ea}
N.~Hama, K.~Hosomichi, and S.~Lee, ``{SUSY Gauge Theories on Squashed
  Three-Spheres},'' \href{http://dx.doi.org/10.1007/JHEP05(2011)014}{{\em JHEP}
  {\bfseries 1105} (2011) 014},
\href{http://arxiv.org/abs/1102.4716}{{\ttfamily arXiv:1102.4716 [hep-th]}}.

\bibitem{Festuccia:2011ws}
G.~Festuccia and N.~Seiberg, ``{Rigid Supersymmetric Theories in Curved
  Superspace},'' \href{http://dx.doi.org/10.1007/JHEP06(2011)114}{{\em JHEP}
  {\bfseries 1106} (2011) 114},
\href{http://arxiv.org/abs/1105.0689}{{\ttfamily arXiv:1105.0689 [hep-th]}}.

\bibitem{Kim:2009wb}
S.~Kim, ``{The Complete superconformal index for N=6 Chern-Simons theory},''
  \href{http://dx.doi.org/10.1016/j.nuclphysb.2012.07.015,
  10.1016/j.nuclphysb.2009.06.025}{{\em Nucl.Phys.} {\bfseries B821} (2009)
  241--284},
\href{http://arxiv.org/abs/0903.4172}{{\ttfamily arXiv:0903.4172 [hep-th]}}.

\bibitem{Imamura:2011su}
Y.~Imamura and S.~Yokoyama, ``{Index for three dimensional superconformal field
  theories with general R-charge assignments},''
  \href{http://dx.doi.org/10.1007/JHEP04(2011)007}{{\em JHEP} {\bfseries 1104}
  (2011) 007},
\href{http://arxiv.org/abs/1101.0557}{{\ttfamily arXiv:1101.0557 [hep-th]}}.

\bibitem{Kallen:2011ny}
J.~Kallen, ``{Cohomological localization of Chern-Simons theory},''
  \href{http://dx.doi.org/10.1007/JHEP08(2011)008}{{\em JHEP} {\bfseries 1108}
  (2011) 008},
\href{http://arxiv.org/abs/1104.5353}{{\ttfamily arXiv:1104.5353 [hep-th]}}.

\bibitem{Ohta:2012ev}
K.~Ohta and Y.~Yoshida, ``{Non-Abelian Localization for Supersymmetric
  Yang-Mills-Chern-Simons Theories on Seifert Manifold},''
  \href{http://dx.doi.org/10.1103/PhysRevD.86.105018}{{\em Phys.Rev.}
  {\bfseries D86} (2012) 105018},
\href{http://arxiv.org/abs/1205.0046}{{\ttfamily arXiv:1205.0046 [hep-th]}}.

\bibitem{Romelsberger:2005eg}
C.~Romelsberger, ``{Counting chiral primaries in N = 1, d=4 superconformal
  field theories},''
  \href{http://dx.doi.org/10.1016/j.nuclphysb.2006.03.037}{{\em Nucl.Phys.}
  {\bfseries B747} (2006) 329--353},
\href{http://arxiv.org/abs/hep-th/0510060}{{\ttfamily arXiv:hep-th/0510060
  [hep-th]}}.

\bibitem{Kinney:2005ej}
J.~Kinney, J.~M. Maldacena, S.~Minwalla, and S.~Raju, ``{An Index for 4
  dimensional super conformal theories},''
  \href{http://dx.doi.org/10.1007/s00220-007-0258-7}{{\em Commun.Math.Phys.}
  {\bfseries 275} (2007) 209--254},
\href{http://arxiv.org/abs/hep-th/0510251}{{\ttfamily arXiv:hep-th/0510251
  [hep-th]}}.

\bibitem{Romelsberger:2007ec}
C.~Romelsberger, ``{Calculating the Superconformal Index and Seiberg
  Duality},''
\href{http://arxiv.org/abs/0707.3702}{{\ttfamily arXiv:0707.3702 [hep-th]}}.

\bibitem{Dolan:2008qi}
F.~Dolan and H.~Osborn, ``{Applications of the Superconformal Index for
  Protected Operators and q-Hypergeometric Identities to N=1 Dual Theories},''
  \href{http://dx.doi.org/10.1016/j.nuclphysb.2009.01.028}{{\em Nucl.Phys.}
  {\bfseries B818} (2009) 137--178},
\href{http://arxiv.org/abs/0801.4947}{{\ttfamily arXiv:0801.4947 [hep-th]}}.

\bibitem{Gadde:2011ia}
A.~Gadde and W.~Yan, ``{Reducing the 4d Index to the $S^3$ Partition
  Function},'' \href{http://dx.doi.org/10.1007/JHEP12(2012)003}{{\em JHEP}
  {\bfseries 1212} (2012) 003},
\href{http://arxiv.org/abs/1104.2592}{{\ttfamily arXiv:1104.2592 [hep-th]}}.

\bibitem{Dolan:2011rp}
F.~Dolan, V.~Spiridonov, and G.~Vartanov, ``{From 4d superconformal indices to
  3d partition functions},''
  \href{http://dx.doi.org/10.1016/j.physletb.2011.09.007}{{\em Phys.Lett.}
  {\bfseries B704} (2011) 234--241},
\href{http://arxiv.org/abs/1104.1787}{{\ttfamily arXiv:1104.1787 [hep-th]}}.

\bibitem{Imamura:2011uw}
Y.~Imamura, ``{Relation between the 4d superconformal index and the $S^3$
  partition function},'' \href{http://dx.doi.org/10.1007/JHEP09(2011)133}{{\em
  JHEP} {\bfseries 1109} (2011) 133},
\href{http://arxiv.org/abs/1104.4482}{{\ttfamily arXiv:1104.4482 [hep-th]}}.

\bibitem{Imamura:2011wg}
Y.~Imamura and D.~Yokoyama, ``{N=2 supersymmetric theories on squashed
  three-sphere},'' \href{http://dx.doi.org/10.1103/PhysRevD.85.025015}{{\em
  Phys.Rev.} {\bfseries D85} (2012) 025015},
\href{http://arxiv.org/abs/1109.4734}{{\ttfamily arXiv:1109.4734 [hep-th]}}.

\bibitem{Drukker:2010jp}
N.~Drukker, D.~Gaiotto, and J.~Gomis, ``{The Virtue of Defects in 4D Gauge
  Theories and 2D CFTs},''
  \href{http://dx.doi.org/10.1007/JHEP06(2011)025}{{\em JHEP} {\bfseries 1106}
  (2011) 025},
\href{http://arxiv.org/abs/1003.1112}{{\ttfamily arXiv:1003.1112 [hep-th]}}.

\bibitem{Hosomichi:2010vh}
K.~Hosomichi, S.~Lee, and J.~Park, ``{AGT on the S-duality Wall},''
  \href{http://dx.doi.org/10.1007/JHEP12(2010)079}{{\em JHEP} {\bfseries 1012}
  (2010) 079},
\href{http://arxiv.org/abs/1009.0340}{{\ttfamily arXiv:1009.0340 [hep-th]}}.

\bibitem{Gaiotto:2009we}
D.~Gaiotto, ``{N=2 dualities},''
  \href{http://dx.doi.org/10.1007/JHEP08(2012)034}{{\em JHEP} {\bfseries 1208}
  (2012) 034},
\href{http://arxiv.org/abs/0904.2715}{{\ttfamily arXiv:0904.2715 [hep-th]}}.

\bibitem{Quine:1993aa}
J.~Quine, S.~Heydari, and R.~Song, ``{Zeta Regularized Products},''
  \href{http://dx.doi.org/10.1090/S0002-9947-1993-1100699-1}{{\em Trans. Amer.
  Math. Soc.} {\bfseries 338} (1993) 213}.

\bibitem{Teschner:2003at}
J.~Teschner, ``{From Liouville theory to the quantum geometry of Riemann
  surfaces},''
\href{http://arxiv.org/abs/hep-th/0308031}{{\ttfamily arXiv:hep-th/0308031
  [hep-th]}}.

\bibitem{Gaiotto:2008sa}
D.~Gaiotto and E.~Witten, ``{Supersymmetric Boundary Conditions in N=4 Super
  Yang-Mills Theory},'' \href{http://dx.doi.org/10.1007/s10955-009-9687-3}{{\em
  J.Statist.Phys.} {\bfseries 135} (2009) 789--855},
\href{http://arxiv.org/abs/0804.2902}{{\ttfamily arXiv:0804.2902 [hep-th]}}.

\bibitem{Gaiotto:2008ak}
D.~Gaiotto and E.~Witten, ``{S-Duality of Boundary Conditions In N=4 Super
  Yang-Mills Theory},'' {\em Adv.Theor.Math.Phys.} {\bfseries 13} (2009) ,
\href{http://arxiv.org/abs/0807.3720}{{\ttfamily arXiv:0807.3720 [hep-th]}}.

\bibitem{Closset:2012ru}
C.~Closset, T.~T. Dumitrescu, G.~Festuccia, and Z.~Komargodski,
  ``{Supersymmetric Field Theories on Three-Manifolds},''
\href{http://arxiv.org/abs/1212.3388}{{\ttfamily arXiv:1212.3388 [hep-th]}}.

\bibitem{Klare:2012gn}
C.~Klare, A.~Tomasiello, and A.~Zaffaroni, ``{Supersymmetry on Curved Spaces
  and Holography},'' \href{http://dx.doi.org/10.1007/JHEP08(2012)061}{{\em
  JHEP} {\bfseries 1208} (2012) 061},
\href{http://arxiv.org/abs/1205.1062}{{\ttfamily arXiv:1205.1062 [hep-th]}}.

\bibitem{Dumitrescu:2012ha}
T.~T. Dumitrescu, G.~Festuccia, and N.~Seiberg, ``{Exploring Curved
  Superspace},'' \href{http://dx.doi.org/10.1007/JHEP08(2012)141}{{\em JHEP}
  {\bfseries 1208} (2012) 141},
\href{http://arxiv.org/abs/1205.1115}{{\ttfamily arXiv:1205.1115 [hep-th]}}.

\bibitem{Alday:2013lba}
L.~F. Alday, D.~Martelli, P.~Richmond, and J.~Sparks, ``{Localization on
  Three-Manifolds},''
\href{http://arxiv.org/abs/1307.6848}{{\ttfamily arXiv:1307.6848 [hep-th]}}.

\bibitem{Gang:2009wy}
D.~Gang, ``{Chern-Simons theory on L(p,q) lens spaces and Localization},''
\href{http://arxiv.org/abs/0912.4664}{{\ttfamily arXiv:0912.4664 [hep-th]}}.

\bibitem{Benini:2011nc}
F.~Benini, T.~Nishioka, and M.~Yamazaki, ``{4d Index to 3d Index and 2d
  TQFT},'' \href{http://dx.doi.org/10.1103/PhysRevD.86.065015}{{\em Phys.Rev.}
  {\bfseries D86} (2012) 065015},
\href{http://arxiv.org/abs/1109.0283}{{\ttfamily arXiv:1109.0283 [hep-th]}}.

\bibitem{Imamura:2012rq}
Y.~Imamura and D.~Yokoyama, ``{${\bf S}^3/\mathbb Z_n$ partition function and
  dualities},'' \href{http://dx.doi.org/10.1007/JHEP11(2012)122}{{\em JHEP}
  {\bfseries 1211} (2012) 122},
\href{http://arxiv.org/abs/1208.1404}{{\ttfamily arXiv:1208.1404 [hep-th]}}.

\bibitem{Hama:2012bg}
N.~Hama and K.~Hosomichi, ``{Seiberg-Witten Theories on Ellipsoids},''
  \href{http://dx.doi.org/10.1007/JHEP09(2012)033,
  10.1007/JHEP10(2012)051}{{\em JHEP} {\bfseries 1209} (2012) 033},
\href{http://arxiv.org/abs/1206.6359}{{\ttfamily arXiv:1206.6359 [hep-th]}}.

\bibitem{Gomis:2012wy}
J.~Gomis and S.~Lee, ``{Exact Kahler Potential from Gauge Theory and Mirror
  Symmetry},''
\href{http://arxiv.org/abs/1210.6022}{{\ttfamily arXiv:1210.6022 [hep-th]}}.

\bibitem{Martelli:2011fu}
D.~Martelli, A.~Passias, and J.~Sparks, ``{The Gravity dual of supersymmetric
  gauge theories on a squashed three-sphere},''
  \href{http://dx.doi.org/10.1016/j.nuclphysb.2012.07.019}{{\em Nucl.Phys.}
  {\bfseries B864} (2012) 840--868},
\href{http://arxiv.org/abs/1110.6400}{{\ttfamily arXiv:1110.6400 [hep-th]}}.

\bibitem{Closset:2013vra}
C.~Closset, T.~T. Dumitrescu, G.~Festuccia, and Z.~Komargodski, ``{The Geometry
  of Supersymmetric Partition Functions},''
  \href{http://dx.doi.org/10.1007/JHEP01(2014)124}{{\em JHEP} {\bfseries 1401}
  (2014) 124},
\href{http://arxiv.org/abs/1309.5876}{{\ttfamily arXiv:1309.5876 [hep-th]}}.

\bibitem{Nian:2013qwa}
J.~Nian, ``{Localization of Supersymmetric Chern-Simons-Matter Theory on a
  Squashed $S^3$ with $SU(2)\times U(1)$ Isometry},''
\href{http://arxiv.org/abs/1309.3266}{{\ttfamily arXiv:1309.3266 [hep-th]}}.

\bibitem{Tanaka:2013dca}
A.~Tanaka, ``{Localization on round sphere revisited},''
  \href{http://dx.doi.org/10.1007/JHEP11(2013)103}{{\em JHEP} {\bfseries 1311}
  (2013) 103},
\href{http://arxiv.org/abs/1309.4992}{{\ttfamily arXiv:1309.4992 [hep-th]}}.

\bibitem{Drukker:2012sr}
N.~Drukker, T.~Okuda, and F.~Passerini, ``{Exact results for vortex loop
  operators in 3d supersymmetric theories},''
\href{http://arxiv.org/abs/1211.3409}{{\ttfamily arXiv:1211.3409 [hep-th]}}.

\bibitem{Witten:1988hf}
E.~Witten, ``{Quantum Field Theory and the Jones Polynomial},''
\href{http://dx.doi.org/10.1007/BF01217730}{{\em Commun.Math.Phys.} {\bfseries
  121} (1989) 351}.

\bibitem{Kao:1995gf}
H.-C. Kao, K.-M. Lee, and T.~Lee, ``{The Chern-Simons coefficient in
  supersymmetric Yang-Mills Chern-Simons theories},''
  \href{http://dx.doi.org/10.1016/0370-2693(96)00119-0}{{\em Phys.Lett.}
  {\bfseries B373} (1996) 94--99},
\href{http://arxiv.org/abs/hep-th/9506170}{{\ttfamily arXiv:hep-th/9506170
  [hep-th]}}.

\bibitem{Tanaka:2012nr}
A.~Tanaka, ``{Comments on knotted 1/2 BPS Wilson loops},''
  \href{http://dx.doi.org/10.1007/JHEP07(2012)097}{{\em JHEP} {\bfseries 1207}
  (2012) 097},
\href{http://arxiv.org/abs/1204.5975}{{\ttfamily arXiv:1204.5975 [hep-th]}}.

\bibitem{Kapustin:2012iw}
A.~Kapustin, B.~Willett, and I.~Yaakov, ``{Exact results for supersymmetric
  abelian vortex loops in 2+1 dimensions},''
\href{http://arxiv.org/abs/1211.2861}{{\ttfamily arXiv:1211.2861 [hep-th]}}.

\end{thebibliography}\endgroup


\vskip-36pt}
\begin{thebibliography}{}

\bibitem[V:1]{Ga} D. Gaiotto,
\href{run:ClassofTheories.pdf}
{``Families of $N=2$ field theories''}
\bibitem[V:2]{N} A. Neitzke,
\href{run:hitchin-systems.pdf}
{``Hitchin systems in ${\mathcal N}=2$ field theory''}
\bibitem[V:3]{T} Y. Tachikawa,
\href{run:Instantons and W-algebras.pdf}
{``A review on instanton counting and W-algebras''}
\bibitem[V:4]{M} K. Maruyoshi,
\href{run:matrix.pdf}
{``$\beta$-deformed matrix models and the 2d/4d correspondence''}
\bibitem[V:5]{P} V. Pestun,
\href{run:S4.pdf}
{``Localization for ${\mathcal N}=2$ Supersymmetric Gauge Theories in Four Dimensions''}
\bibitem[V:6]{O} T. Okuda,
\href{run:line-operators.pdf}
{``Line operators in supersymmetric gauge theories and the 2d-4d relation''}
\bibitem[V:7]{Gu} S. Gukov,
\href{run:surface/surface.pdf}
{``Surface Operators''}
\bibitem[V:8]{RR} L. Rastelli, S. Razamat,
\href{run:N=2 index review/reviewindex2.pdf}
{``Index of theories of class $\mathcal S$: a review''}
\bibitem[V:9]{H} K. Hosomichi,
\href{run:s3 partition function.pdf}
{``A review on SUSY gauge theories on $S^3$''}
\bibitem[V:10]{D} T. Dimofte,
\href{run:3d3d.pdf}
{``3d Superconformal Theories from Three-Manifolds''}
\bibitem[V:11]{T} J. Teschner,
\href{run:teschner.pdf}
{``Supersymmetric gauge theories, quantization of ${\mathcal M}_{\mathrm flat}$, and Liouville theory''}
\bibitem[V:12]{A} M. Aganagic and S. Shakirov,
\href{run:triality.pdf}
{``Gauge/vortex duality and AGT''}
\bibitem[V:13]{KW} D. Krefl, J. Walcher,
\href{run:Bmodel.pdf}
{``B-Model Approaches to Instanton Counting''}
\end{thebibliography}

\paragraph{\large References to articles in this volume}
\renewcommand{\refname}{

\end{document}